\documentclass[conference]{IEEEtran}
\usepackage{cite}
\usepackage{amsmath,amssymb,amsfonts}
\usepackage{algorithmic}
\usepackage{graphicx}
\usepackage{textcomp}
\usepackage{xcolor}
\usepackage[hyphens]{url}

\usepackage{multirow}
\usepackage{mathtools}
\usepackage[colorlinks=true,linkcolor=blue,citecolor=magenta]{hyperref}

\def\BibTeX{{\rm B\kern-.05em{\sc i\kern-.025em b}\kern-.08em
    T\kern-.1667em\lower.7ex\hbox{E}\kern-.125emX}}

\newcommand{\DesName}{DisaggRec}

\newif\ifrev
\ifrev
  \newcommand{\note}[1]{{\color{red} [#1]}}
  \newcommand{\revise}[1]{{\color{blue} #1}}
\else
  \newcommand{\note}[1]{}
  \newcommand{\revise}[1]{{\color{black} #1}}
\fi

\title{{\DesName}: Architecting Disaggregated Systems for Large-Scale Personalized Recommendation}
\author{\normalsize 
Liu Ke${^\dagger}{^\ddagger}{^\ast}$,
Xuan Zhang$^\ddagger$,
Benjamin Lee${^\dagger}{^\diamond}$,
G. Edward Suh${^\dagger}{^\S}$, 
Hsien-Hsin S. Lee${^\dagger}{^\P}{^\ast}$
\\
\normalsize
$^\dagger$Meta AI,
$^\ddagger$Washington University in St. Louis,
$^\diamond$University of Pennsylvania,
$^\S$Cornell University,
$^\P$Intel

} 
\begin{document}
\maketitle
\thispagestyle{plain}
\pagestyle{plain}

{\let\thefootnote\relax\footnotetext{$^\ast$ Work done while at Meta AI.}}

%%%%%% -- PAPER CONTENT STARTS-- %%%%%%%%

\begin{abstract}

Deep learning-based personalized recommendation systems are widely used for online user-facing services in production datacenters, where a large amount of hardware resources are procured and managed to reliably provide low-latency services without disruption.
As the recommendation models continue to evolve and grow in size, our analysis projects that datacenters deployed with monolithic servers will spend up to 12.4$\times$ total cost of ownership (TCO) to meet the requirement of model size and complexity over the next three years.
Moreover, through in-depth characterization, we reveal that the monolithic server-based cluster suffers resource idleness and wastes up to 30\% TCO by provisioning resources in fixed proportions.
To address this challenge, we propose {\DesName}, a disaggregated system for large-scale recommendation serving.
{\DesName} achieves the independent decoupled scaling-out of the compute and memory resources to match the changing demands from fast-evolving workloads. It also improves system reliability by segregating the failures of compute nodes and memory nodes.
These two main benefits from disaggregation collectively reduce the TCO by up to 49.3\%.
Furthermore, disaggregation enables flexible and agile provisioning of increasing hardware heterogeneity in future datacenters.
By deploying new hardware featuring near-memory processing capability, our evaluation shows that the disaggregated cluster achieves 21\%$\sim$43.6\% TCO savings over the monolithic server-based cluster across a three-year span of model evolution.

\end{abstract}

\section{Introduction}
\label{sec:introduction}

Personalized recommendation is a fundamental service for many Internet services such as search engines, social networks, online retailing, and content streaming~\cite{GCP,Walmart_AI,Amazon_Personalize,alibabaRec,youtube}, all of which are now tightly interwoven into our daily lives.
These online services are predominantly powered by recommendation models that leverage modern deep learning technologies to achieve high prediction accuracy and delivery quality individualized user experiences~\cite{hazelwood2018applied}. Following recent algorithmic innovations~\cite{wnd,mtwnd,DLRM,din,dien,rec-training-ZionEX}, recommendation models are expected to grow in size at an accelerated pace in order to keep up with the rapidly increasing and evolving data features. Therefore, a large quantity of datacenter hardware serves recommendation queries. This hardware often consists of monolithic servers as basic building blocks, configured with a mix of CPUs, GPUs, and memory (DRAM).

However, as recommendation models continue to evolve and scale up, monolithic servers will face two aggravating challenges. First, these models require tremendous memory capacity that will soon exceed what can be feasibly integrated within a server. Second, model variants will require compute and memory of dissimilar proportions and intensities. Embedding operations will stress memory whereas dense operations are computation-demanding. Because the mix of these operations will continue to evolve and vary across models, no one-size-fits-all design for monolithic servers can simultaneously optimize the cost and energy efficiency across all the model variants.

Resource disaggregation decouples the deployment of compute and memory, allowing system architects to provision and scale resource types independently. Compute nodes (CNs) supply high-performance processors but only a limited amount of memory, whereas memory nodes (MNs) supply many high-capacity DRAM devices. MNs can integrate small processors for general-purpose compute \cite{farmem-nsdi-2014,disaggmem-nsdi-2017,legoos,disaggmem-eurosys-2018,farmem-eurosys-2020,farmem-osdi-2020,disaggmem-osdi-2020}, application-specific accelerators \cite{farview,clio}, or no processing beyond the minimum needed for control and communication \cite{hp-themachine,hp-memory-driven-computing,disaggmem-atc-2020,disaggmem-atc-2021}. In disaggregated systems, an application receives networked allocations of CNs and MNs that can best match its needs.
\revise{Previous work has shown disaggregation can benefit resource utilization by up to 50\%~\cite{legoos}, improve system reliability by 17\%--49\%~\cite{legoos}, and reduce hardware cost by 7\%~\cite{Pond}. However, performance may also suffer at the same time when deployed for large-scale production workloads. LegoOS~\cite{legoos} showed 34\%--68\% performance degradation, and Pond~\cite{Pond} reported a performance loss of more than 25\%.}

In this paper, we propose {\DesName} to address infrastructure-level challenges from fast-evolving recommendation systems. {\DesName} is a disaggregated system for recommendation serving at scale. The system must meet strict performance and availability targets for the recommendation system is interactive and user-facing. For performance, we characterize task scheduling and optimize disaggregated system configurations to meet the target tail-latency required by the service-level agreement (SLA). For availability, we characterize task loads and component failures, and dynamically provision compute and memory resources to ensure reliable service under occasional system faults. Collectively, our work makes the following contributions. % significant contributions. 

\textbf{Quantifying Inefficiency of Monolithic Servers.} First, we characterize deployed production-grade recommendation systems comparing scale-up and scale-out strategies using monolithic servers. We find monolithic design is inefficient in terms of the total cost of ownership (TCO). Monolithic servers provision compute and memory in fixed proportions, leading to idle resources and wasted costs of up to 23.1\%. Furthermore, they must guard against server failures, resulting in over-provisioned resources and wasted costs of up to 6.8\%.

\textbf{Improving Efficiency with Disaggregation.} Second, we co-optimize the partitioning strategies for recommendation models and design strategies for disaggregated CNs and MNs. We minimize the cost subject to latency targets and availability requirements for recommendation queries. {\DesName} \revise{experiences minor throughput degradation ($<$2\%) and} reduces cost by 49.3\% when compared to a monolithic design.

\revise{
%\textbf{Managing Heterogeneity with Disaggregation.}
\textbf{Provisioning Heterogeneity with Disaggregation.}
Hardware accelerators are increasingly deployed in production datacenters to optimize operational efficiency. In addition to compute-centric accelerators for dense computation, near-memory processing DRAM is reaching the market to help accelerate workflows bounded by memory capacity and memory bandwidth. However, monolithic servers are inept and far less flexible in adopting emerging hardware because their tightly bundled hardware components preclude workloads from receiving their most cost-effective resource mixes. The disaggregated cluster addresses these challenges, reducing cost by 21\% to 43.6\% compared to the monolithic server-based cluster.
}
%In contrast, disaggregated servers facilitate the agile deployment of diverse hardware by provisioning them in separate resource pools and allowing each workload to receive their optimal mix. Overall, disaggregated design achieves 27.2\% to 43.4\% TCO saving over monolithic design.

\section{Recommendation Systems}
\label{sec:background}

\begin{figure}[t!]
  \centering
  \includegraphics[width=\columnwidth]{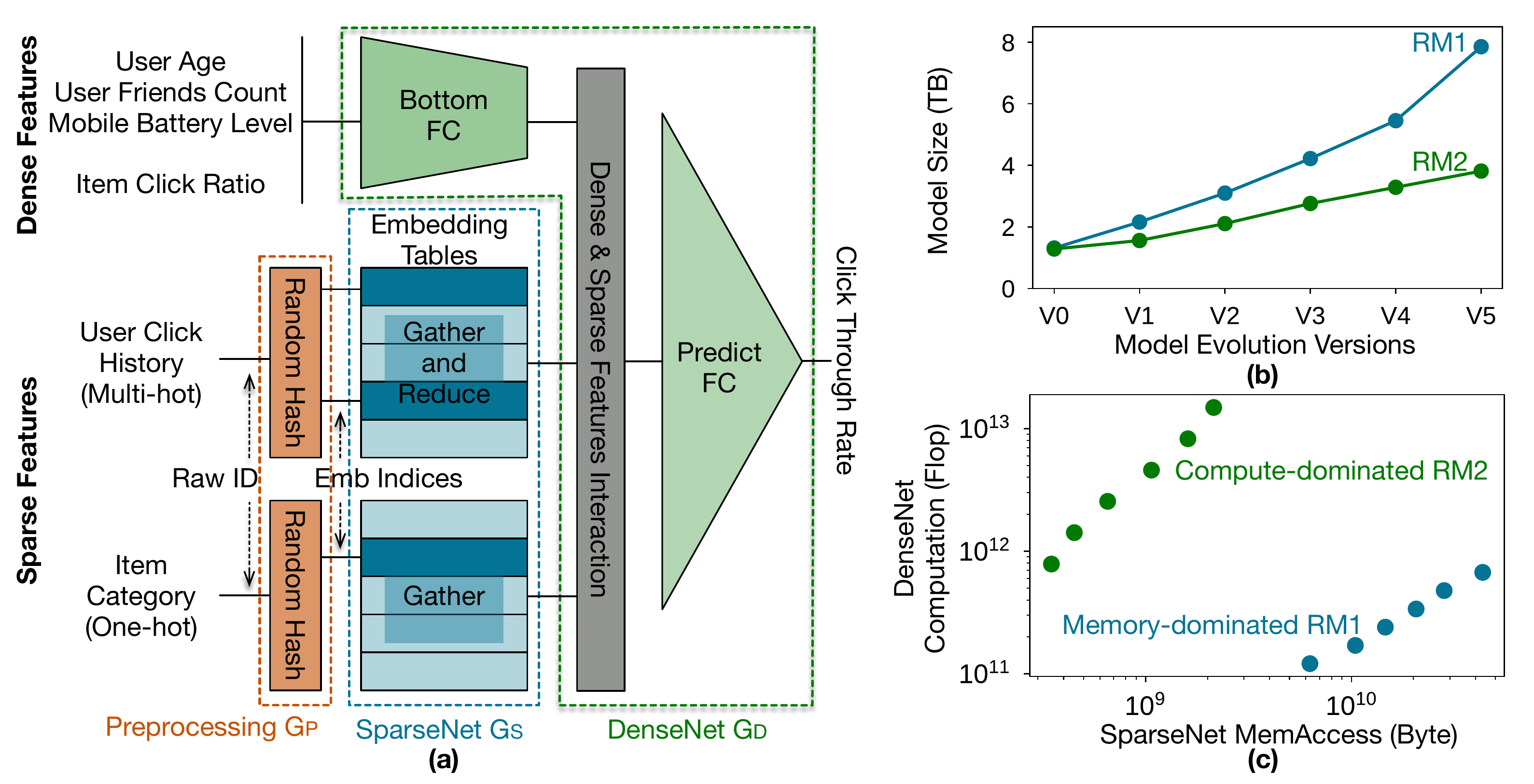}
  \vspace{-0.8cm}
  \caption{(a) General recommendation model; 
  Synthetic evolution projection of (b) model size and (c) model complexity for memory-dominated RM1 and compute-dominated RM2 models.}
  \label{fig:model.projection}
  \vspace{-0.5cm}
\end{figure}

Personalized recommendation models are widely deployed by Internet service providers to enhance user experiences, contributing more than 80\% of AI cycles in modern hyperscale datacenters~\cite{gupta-hpca-2020}. Figure~\ref{fig:model.projection}(a) presents the three major computational components of a recommendation model. Pre-processing ($G_P$) employs hash functions to map input signals of raw, sparse features ({\em e.g.}, user ID, webpage ID) to corresponding indices in the embedding tables. SparseNet ($G_S$) performs sparse, irregular memory lookups with the embedding tables as well as pooling operations. DenseNet ($G_D$) evaluates compute-intensive, fully-connected (FC) layers.

\textbf{Model Scaling.} Recommendation models continue to evolve and scale rapidly, increasing both in size and complexity. In this paper, we focus on two industry-grade models---a memory-intensive RM1 and a compute-intensive RM2. Moreover, in Figure~\ref{fig:model.projection}(b), we use internal projections to estimate their scaling trends over six model generations that span the next three years. For RM1, SparseNet is the primary growth driver, increasing the model size from 1.4TB to 7.8TB and making memory the main resource bottleneck. In contrast, for RM2, DenseNet is the primary growth driver, increasing the depth and width of FC layers, which in turn increases the number of FLOPs by 18.9$\times$ and makes compute the main bottleneck.

\textbf{Service Requirements.} The user-facing recommendation services must provide each query a timely and accurate response. Billions of users around the world expect high availability from these services. 
Thus, design and management strategies for recommendation systems must navigate complex interactions among latency, throughput, and resilience. 

\begin{figure}[t!]
  \centering
  \includegraphics[width=\columnwidth]{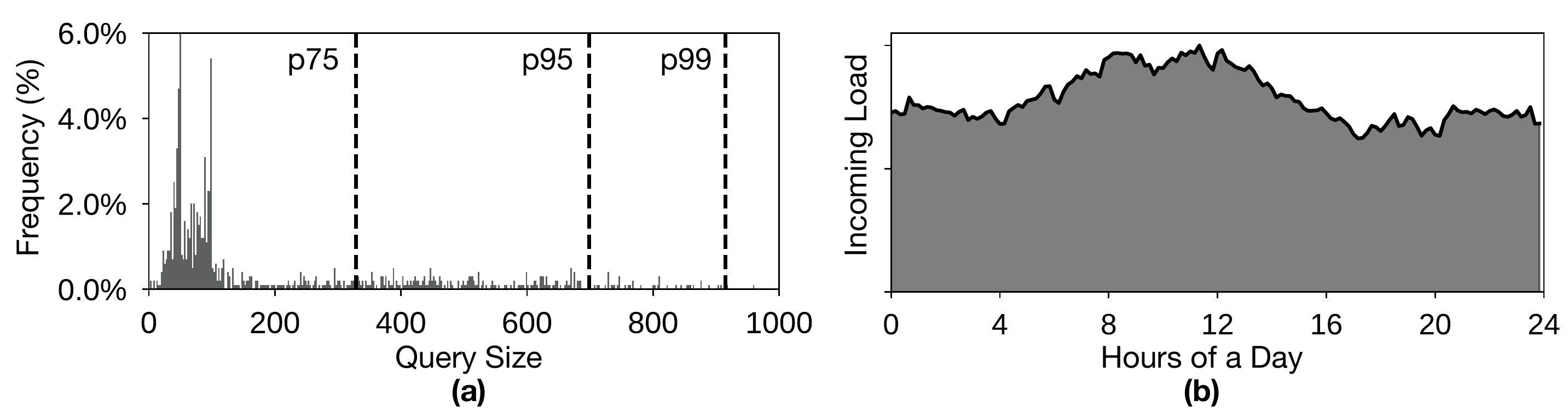}
  \vspace{-0.8cm}
  \caption{(a) Query size distribution of recommendation inference;
  (b) Diurnal load of recommendation services during a day.}
  \label{fig:query.pattern}
  \vspace{-0.6cm}
\end{figure}

Recommendation systems are required to satisfy the service-level agreement (SLA), a contract between the service provider and the end user.
Here, we consider two SLAs for tail-latency and availability.
First, the SLA may require the 95th-percentile (p95) latency to be within 100ms. Achieving these performance goals is complicated due to the dynamic query arrival patterns~\cite{DeepRecSys,Hercules}. Figure~\ref{fig:query.pattern}(a) shows the heavy-tailed distribution of incoming query sizes. 
Second, recommendation systems are deployed on a large fleet of servers to guarantee service availability. Figure~\ref{fig:query.pattern}(b) shows the diurnal loads of recommendation service during a day. Resource allocations tend to be over-provisioned, guarding against machine failures but producing large gaps between typical and peak loads.

%, and the over-provisioning capacity as backups for machine failures. 

\textbf{Production Recommendation Systems.}
\revise{
Production datacenters often customize server design for high-performance recommendation systems to better support a large volume of business-critical services. For example, Meta's Zion/ZionEX server is tailored to support at-scale recommendation~\cite{zion,rec-training-ZionEX}, integrating abundant resources within a machine (e.g., four CPU sockets, 1.5TB of DRAM, eight NVIDIA A100 GPUs each with 40GB of HBM). To meet workloads' increasing demands for computational resources, industry production systems usually follow \textit{scale-up} or \textit{scale-out} strategies.
}

%Nonetheless, even the high-end ZionEX is insufficient for exponentially increasing demands from fast-evolving recommendation workloads.

A scale-up strategy deploys additional resources within a single server.
For example, the scale-up dual-socket (SU-2S) server (detailed configurations in Table~\ref{tab:sys_config}) is a next-generation platform for larger recommendation models that scales up to use two CPU sockets with 2TB of DRAM and eight A100 GPUs, each with 80GB of HBM.

On the other hand, a scale-out strategy distributes a workload across multiple servers as recommendation models will invariably exhaust the limited resources that can be accommodated by a single server.
\revise{
\textit{Distributed inference}~\cite{lui-iiswc-2020} launches a single large model across a group of servers. Their collective memory capacity holds model parameters. The distributed serving paradigm splits the model into computational sub-graphs, called \textit{shards}, and uses remote procedure calls (RPCs) for the execution of sub-graphs and the communication between shards~\cite{pytorch}.
} Sharding permits simpler, smaller servers. For example, the scale-out single-socket (SO-1S) server comprises only one processor socket and 1TB of DRAM with varying 1, 2, 4 GPUs configurations to serve diverse workloads with varied memory and compute intensities (Table~\ref{tab:sys_config}).

\revise{
Optimizing the combination of hardware components for next-generation datacenter servers is increasingly difficult. Domain-specific accelerators and technologies are emerging as component options for system architects. Recommendation model's DenseNet operations might benefit from compute-centric accelerators such as NVIDIA GPUs, Google TPUs~\cite{tpu}, and Alibaba's Hanguang~\cite{hanguang}. While the embedding operations in recommendation models might benefit from near-memory processing (NMP) and processing-in-memory (PIM)~\cite{AxDIMM,samsung-hbmpim,samsung-aquabolt,upmem,hynix-gddr-aim}.}

% are slated to join the future datacenter fleet.

In summary, architects must take a holistic view of system design. 
As recommendation models evolve and hardware devices diversify, system architects must perform their due diligence prior to procurement and deployment. They must carefully evaluate and identify an efficient choice of server components, balancing scale-up and scale-out strategies. In this paper, we will quantify inefficiencies from monolithic servers and pinpoint opportunities for hardware disaggregation to tackle these system challenges.

\section{Model Inference on Monolithic Servers}
\label{sec:insights}

\revise{
We thoroughly characterize industry-grade recommendation workloads on monolithic servers following scale-up and scale-out strategies, producing two key insights.
\begin{itemize}
    \item Embedding reduction performed inside SparseNet shards can greatly reduce the amount of data that need to be communicated between local and remote shards; embedding accesses and reduction should be done using local memory to eliminate unbalanced memory utilization in NUMA or unnecessary remote communication in scale-out systems.
    %\item Embedding reduction performed inside SparseNet shards eliminates NUMA's unbalanced memory utilization and minimizes local-remote communication traffic.
    
    \item \revise{Modern network bandwidth ($\sim$25 GB/s) is comparable to processor interconnects ($\sim$55 GB/s), enabling better scale-out systems as recommendation models scale in size and complexity.
    %scaling-out incurs only $<$5\% performance degradation compared to scaling-up for recommendation inference
    }
    %\item \revise{Modern network bandwidth ($\sim$25 GB/s) is comparable to processor interconnects ($\sim$55 GB/s), which incurs only $<$5\% performance degradation for recommendation inference, and enables better scale-out systems as recommendation models scale in size and complexity.}
    
\end{itemize}
These insights guide DisaggRec's design choices later in Sec~\ref{sec:design}.
All monolithic server configurations used in the following sections are listed in Table~\ref{tab:sys_config}.
}

\begin{figure*}[t!]
  \vspace{-0.1cm}
  \centering
  \includegraphics[width=\textwidth]{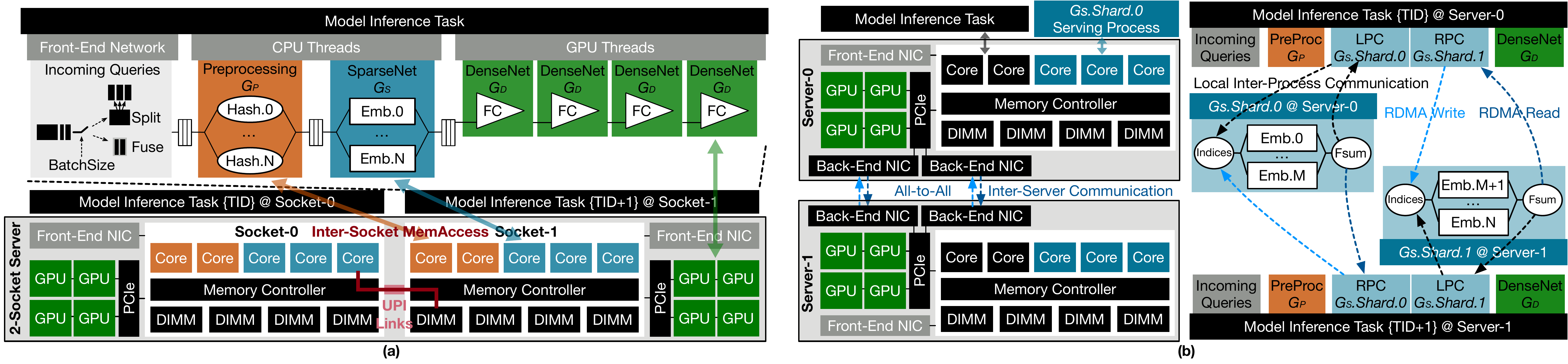}
  \vspace{-0.85cm}
  \caption{(a) Scaling Up: end-to-end model inference on a scale-up (SU-2S) server;
  (b) Scaling Out: distributed inference on two scale-out (SO-1S) servers}
  \label{fig:model.inference}
  \vspace{-0.6cm}
\end{figure*}

\subsection{Scaling Up -- Inference on a Single Server}
A scale-up strategy aims to equip a single server node with sufficient resources to serve end-to-end model inference. 
\revise{
Figure~\ref{fig:model.inference}(a) shows the end-to-end model inference configuration on a scale-up dual-socket (SU-2S) server.
Two inference tasks are launched in parallel on the two processors. 
}
The two front-end network interface cards (NICs) receive incoming queries and return prediction outcomes. Given the dynamic query arrival pattern and the configured batch size, a large query is split into multiple sub-batches and multiple small queries are fused into one large batch. These batches are queued and wait for execution. The inference task exploits three types of parallelism.

\textbf{Model Parallelism.} 
The model's computation graph $G_M$ is partitioned into three sub-graphs, preprocessing $G_P$, SparseNet with  embedding tables $G_S$, and DenseNet $G_D$. Sub-graphs are launched concurrently and pipelined. $G_P$ and $G_S$ co-locate on the CPU's 40 cores while $G_D$ runs on the GPUs within a SU-2S server.

%an Icelake CPU with 40 cores 

\textbf{Operator Parallelism.} 
Physical cores are statically assigned to one single thread where independent operators inside the computation graph can be executed in parallel. 
On a SU-2S server, we assign 20 CPU cores to the preprocessing thread and 20 to the SparseNet thread such that $G_P$'s random hash operators and $G_S$'s embedding operators are executed in parallel on their assigned CPU cores.

\textbf{Data Parallelism.} 
Query batches are executed in parallel across the GPUs. Each GPU launches a replica of DenseNet's threads. On a SU-2S, a processor socket is connected to four A100 GPUs and four $G_D$ threads serve batches of intermediate results from $G_S$.

% in this case, the embeddings expand the two sockets
% therefore ..
% this could lead to unbalanced

We observe degraded performance due to non-uniform memory accesses (NUMA) when SparseNet exceeds the memory capacity of a single socket. As embeddings occupy both memory nodes, the two SparseNet threads on the two processors route half of its accesses to local memory and the other half to remote memory via processor socket interconnect---Intel's Ultra Path Interconnect (UPI)---leading to unbalanced memory bandwidth utilization. Inter-socket bandwidth ($\sim$55 GB/s at peak) is much lower than local memory bandwidth ($\sim$145 GB/s at peak). 

%So, the overall performance is degraded and constrained by the inter-socket memory accesses.

% Connect to result: We later characterized the ...

\subsection{Scaling Out -- Inference on Multiple Servers}

Following the scale-out strategy, the model's SparseNet is sharded and distributed across multiple servers when the embedding tables cannot fit into a single server's memory.
The model's computational graph $G_M$ is partitioned into three sub-graphs for preprocessing ($G_P$), SparseNet shards ($G_S.Shard.0$, $G_S.Shard.1$), and DenseNet ($G_D$).
%The embedding table is the basic sharding unit, e.g. $embedding_{[0..M]}$ in $G_S.Shard.0$ and $embedding_{[(M+1)..N]}$ in $G_S.Shard.1$.

Figure~\ref{fig:model.inference}(b) shows the \textit{distributed inference}~\cite{lui-iiswc-2020} across two scale-out single-socket (SO-1S) servers.
Each server launches two processes, the primary inference task and one SparseNet shard. 
The primary task receives incoming queries, performs preprocessing, routes SparseNet's embedding lookups to the appropriate shards, receives the final summation (Fsum) back from the remote shards, and then executes DenseNet to get the final prediction.
The embedding operations in the primary task are issued by remote procedure calls (RPCs) using pre-stored destination metadata ({\em e.g.}, model ID, SparseNet shard ID).
Embedding operations in one SparseNet shard is packed in one packet which is routed to that shard's serving process hosting targeted SparseNet partitions.
The dedicated back-end network connects servers' back-end NICs using RDMAs, permitting efficient communication between servers.

\subsection{Comparison of Scaling Up and Scaling Out}

\begin{figure}[t!]
  \centering
  \includegraphics[width=\columnwidth]{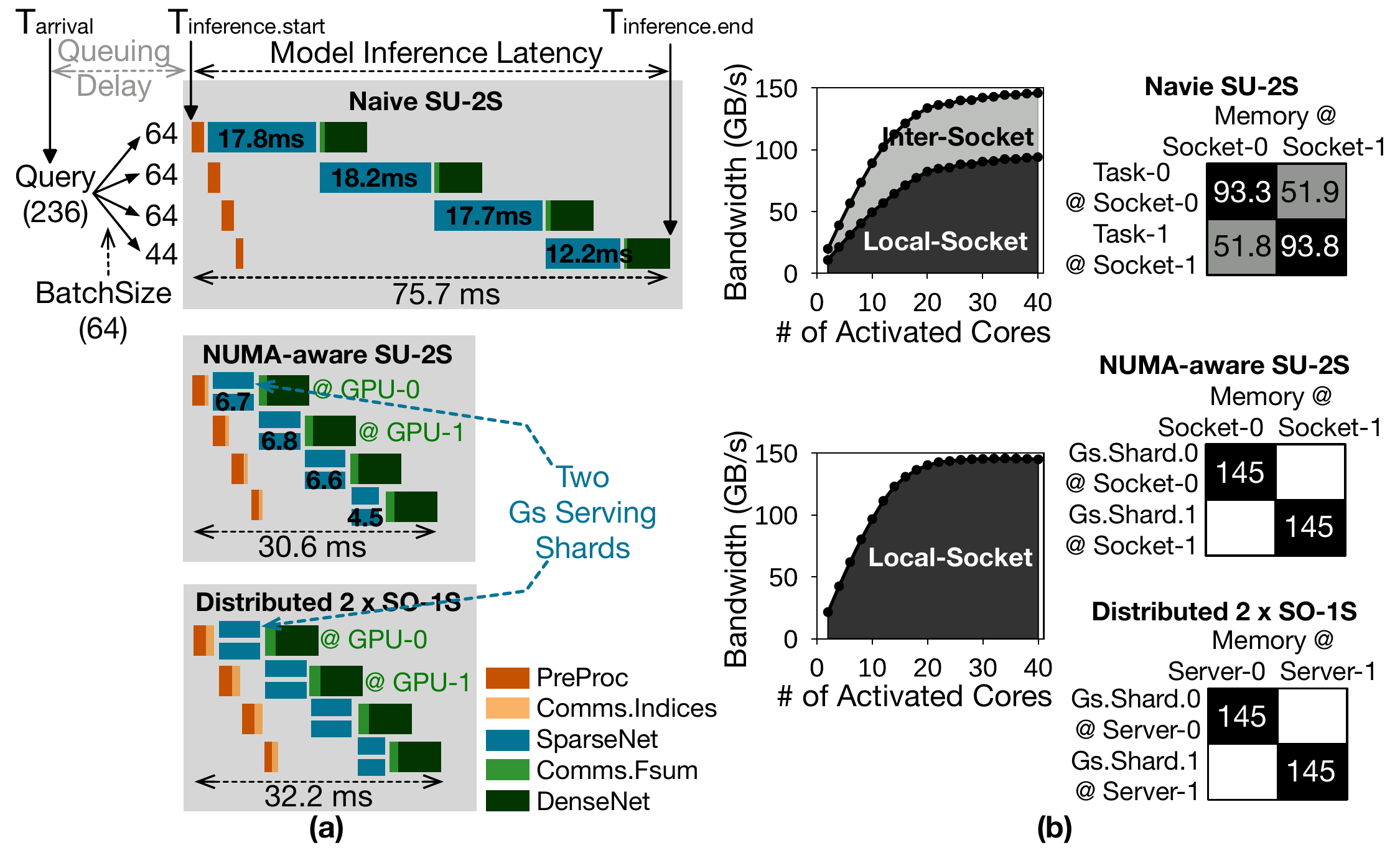}
  \vspace{-0.8cm}
  \caption{(a) Model inference of RM1.V0 on the a single SU-2S server and two distributed SO-1S servers;
  (b) Measured memory bandwidth utilization (GB/s) in the three configurations.}
  \label{fig:pipelines}
  \vspace{-0.6cm}
\end{figure}

To compare the scaling up and scaling out, we perform the end-to-end model inference on a scale-up dual-socket server (SU-2S) with 8 GPUs, and distributed inference on a group of scale-out single-socket servers (SO-1S) with 4 GPUs.

\textbf{Latency.}
Figure~\ref{fig:pipelines} shows the lifecycle of an incoming query, queuing delay and model inference pipelines.
We find the model inference on the SU-2S server suffers from unbalanced local-remote sockets' memory bandwidth utilization. 
In Figure~\ref{fig:pipelines}(b), half of the memory accesses are routed to the local socket's memory (93 GB/s at peak), and the other half of memory accesses are routed to the remote socket's memory (only 52 GB/s at peak) that is bounded by UPI links.

To eliminate memory accesses through UPI interconnect, we implement NUMA-aware inference on the SU-2S adopting the SparseNet sharding scheme to perform memory operations inside local sockets.
Similar to distributed inference, this optimization shards SparseNet $G_s$ into $G_s.Shard.0$ and $G_s.Shard.1$.
Two shards are separately launched on the two processor sockets in parallel.
All memory accesses are routed to the processor socket's local memory. The embedding reduction is performed inside SparseNet shards before sending them to the remote socket.
In Figure~\ref{fig:pipelines}(a), eliminating NUMA's effect within the SU-2S server reduces SparseNet execution time by more than 60\%.
Moreover, the communication overheads are minimal ($<$8\%) that only required for the embeddings' input (lookup indices) and output (Fsum).

The major difference between NUMA-aware inference and distributed inference is the communication interface, UPI links vs NICs.
Today's network communication bandwidth is approaching that of the processor interconnect.
In Figure~\ref{fig:pipelines}(a), the model inference latency of distributed inference on two distributed SO-1S servers only has a minor increment over the NUMA-aware inference on one SU-2S server.
The network bandwidth between SO-1S servers achieves $\sim$25 GB/s at peak, around half of UPI bandwidth,
\revise{incurring less than 5\% performance degradation for scale-out.}
One single server will finally meet the difficulty of holding enough resources to serve the model, (\textit{e.g.}, power delivery of one rack).
Thus, distributed inference provides higher scalability and is a better strategy to support future model growth.

%\textit{Insight-2: today's fast network communication enables better scalability in the scale-out distributed systems, particularly important as recommendation models rapidly scale in size and complexity.}

\begin{figure}[t!]
  \centering
  \includegraphics[width=\columnwidth]{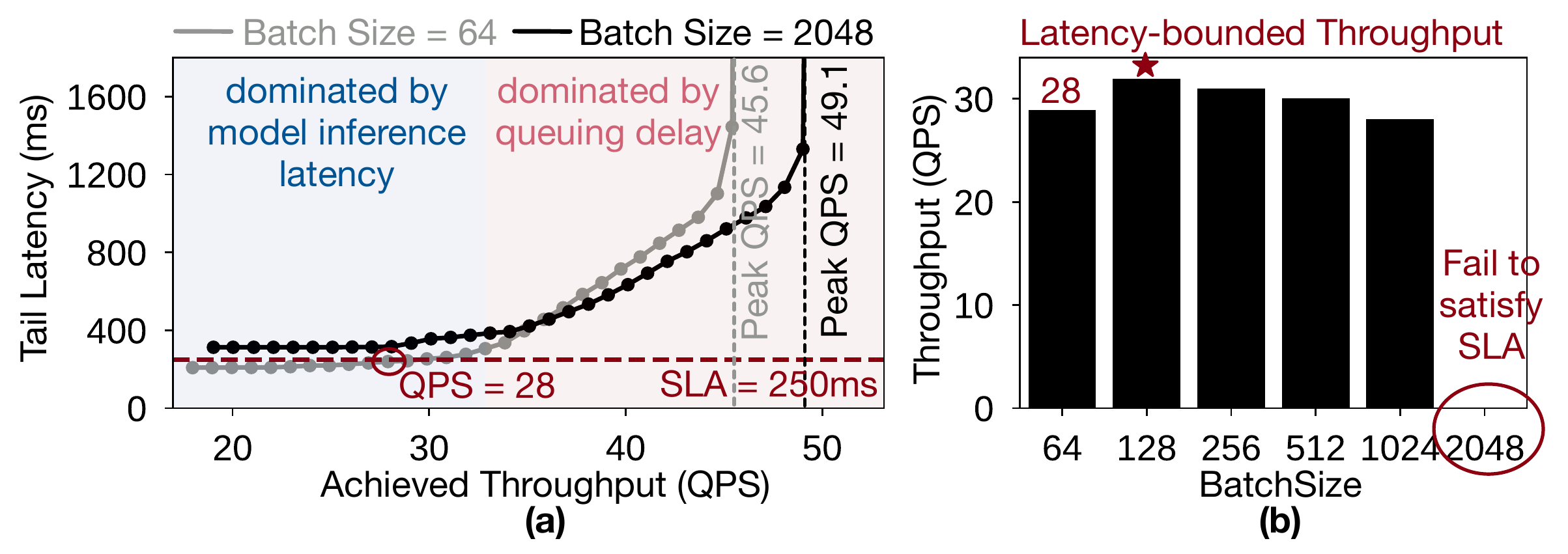}
  \vspace{-0.88cm}
  \caption{(a) Throughput-latency tradeoff and (b) latency-bounded throughput of RM1.V0 launched on two SO-1S servers.}
  \label{fig:SLA.qps}
  \vspace{-0.6cm}
\end{figure}

\textbf{Throughput.}
Unlike datacenter batch workloads that only maximize throughput, the recommendation system maximizes throughput subject to a strict latency SLA.
During workload initialization, we use a hill-climbing algorithm and a pressure test that sweeps query arrival rates and batch sizes \cite{DeepRecSys}. The algorithm halts when latency-bounded throughput plateaus or decreases.
Figure~\ref{fig:SLA.qps}(a) shows that the end-to-end latency increases and gets dominated by queuing delay when query arrival rates are high.
Figure~\ref{fig:SLA.qps}(b) indicates that a batch size of 128 maximizes latency-bounded throughput and further increasing batch size harms throughput. Target latency cannot be met when batch size is 2048 as latency violates the SLA. 

% this method allow us evaluate latency-bounded throughput

\section{{\DesName} System Design}
\label{sec:design}

%Note: memory node is application-specific for recommendation
%Distributed inference: shared memory accesses to load all embeddings --> synchronous

\note{inefficiency number \& refer to evaluation section number}
\note{cite legoos for the reasons of disaggregation, conceptually}

\revise{Given diverse, evolving characteristics of recommendation workloads, tightly coupling the provision of compute and memory in a monolithic server produces up to 30\% wasted cost on idle resources (see Sec~\ref{sec:inefficiency}). 
Resource disaggregation~\cite{legoos} permits independent resource scaling, better failure isolation, and flexible heterogeneity provisioning.
We propose {\DesName} to optimize the total cost of ownership when serving large-scale recommendation systems.

At a glance, memory disaggregation may seem like a bad idea for large recommendation models, whose performance is known to be often limited by memory bandwidth. Yet, we show that disaggregated architecture can be an effective solution when embeddeding reduction can be done locally. 
}

\vspace{-0.1cm}
\subsection{{\DesName} Overview}
\begin{figure}[t!]
  \centering
  \includegraphics[width=\columnwidth]{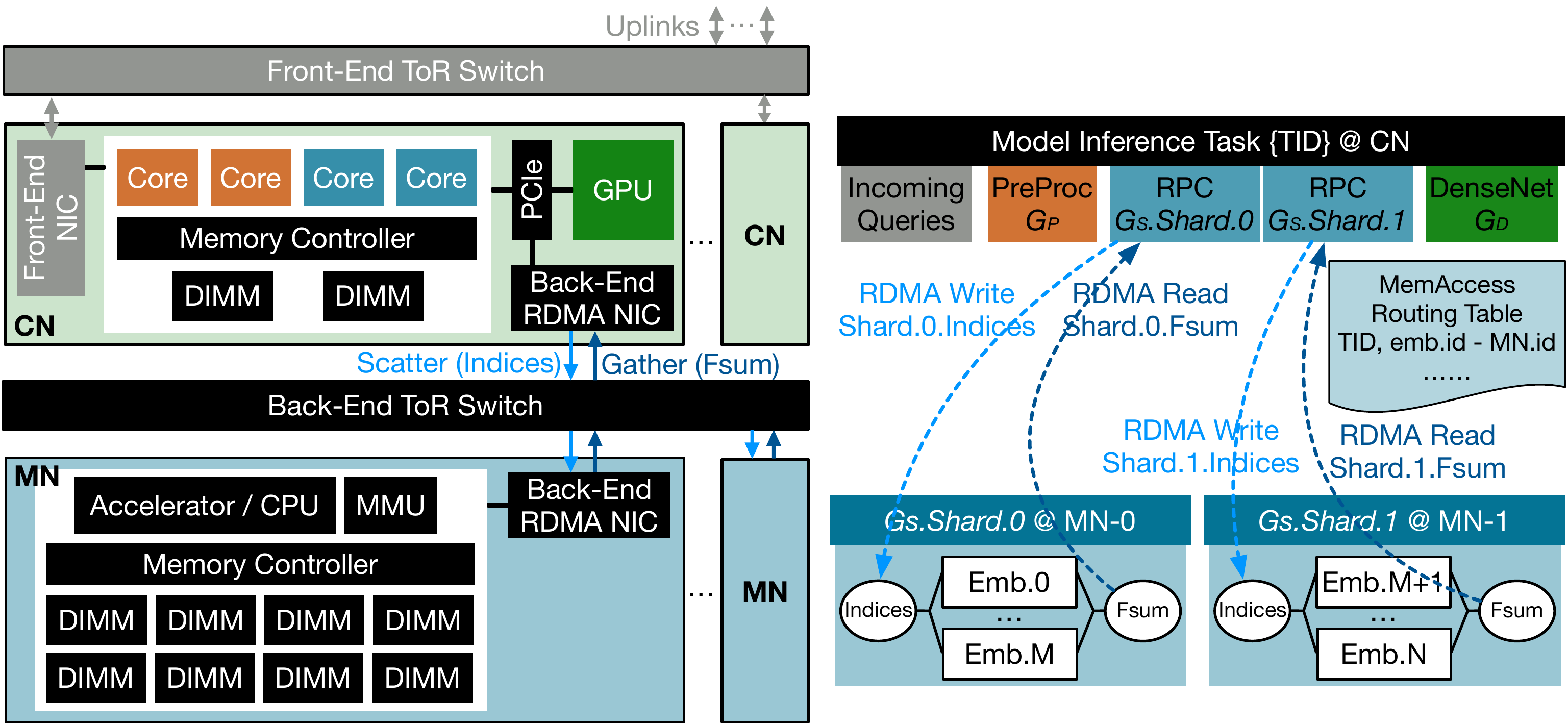}
  \vspace{-0.75cm}
  \caption{Disaggregated system architecture and RPC-based model serving.}
  \label{fig:disagg.system}
  \vspace{-0.4cm}
\end{figure}

\textbf{System Architecture.}
\note{Emphasize the reduction in memory nodes}
Figure~\ref{fig:disagg.system} describes how hardware resources in traditional monolithic servers are disseminated into network-attached compute nodes (CNs) and memory nodes (MNs). This architecture resembles those in prior work~\cite{hp-themachine,legoos,farview}. But whereas prior MN designs offer transparent physical memory with no processing power ~\cite{hp-themachine, disaggmem-atc-2020, disaggmem-atc-2021}, {\DesName}'s MN includes an ASIC or a light-weight processor to perform embedding reduction locally. % with abundant memory. 

The decision to process near memory arises from our insights in Sec ~\ref{sec:insights}. The embedding reduction performed inside the remote SparseNet shards can reduce communication traffic by transferring only the embeddings' input and output values (indices, Fsum). Without such processing, the recommendation system would access raw embedding entries at the remote MNs and incur significant network overheads. 

%(from hundreds of GBs to several TBs).

%To enable fast network communication between CNs and MNs, the 

An RDMA-supported network topology connects all CNs and MNs together. Every node has a dedicated back-end NIC connected with the back-end ToR switch. The fast network enables low-latency and high-bandwidth communication, leading to only minor latency overhead when transferring embeddings' indices and Fsum.

\begin{figure}[t!]
  \centering
  \includegraphics[width=\columnwidth]{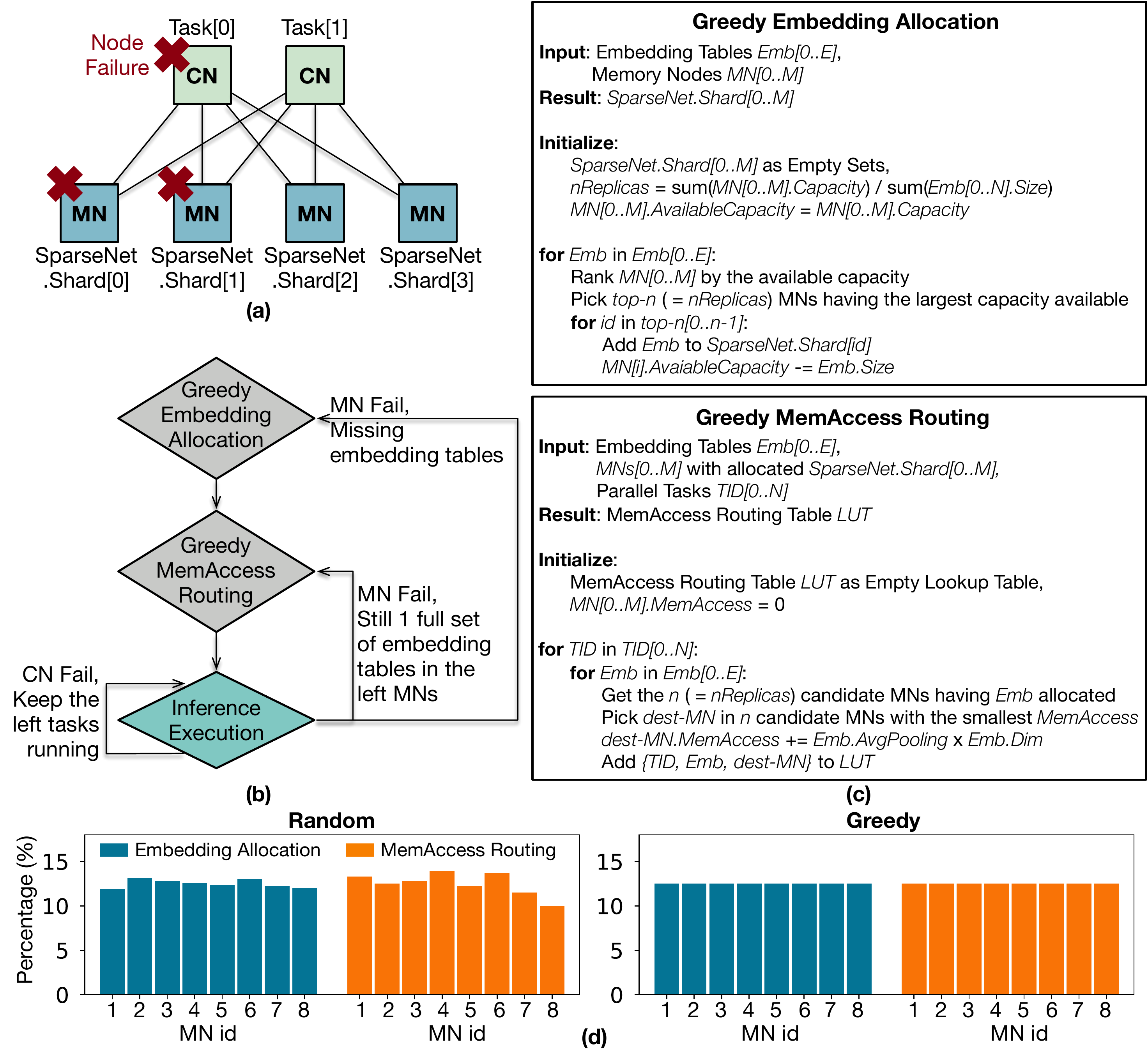}
  \vspace{-0.85cm}
  \caption{(a) One model inference serving unit with 2 CNs and 4 MNs;
  (b) The processes to handle CN and MN failures;
  (c) Greedy embedding allocation and MemAccess routing methods;
  (d) Comparison of naive random and our greedy embedding management.} 
  \label{fig:process.memory.managment}
  \vspace{-0.5cm}
\end{figure}

\begin{figure*}[t!]
  \centering
  \includegraphics[width=\textwidth]{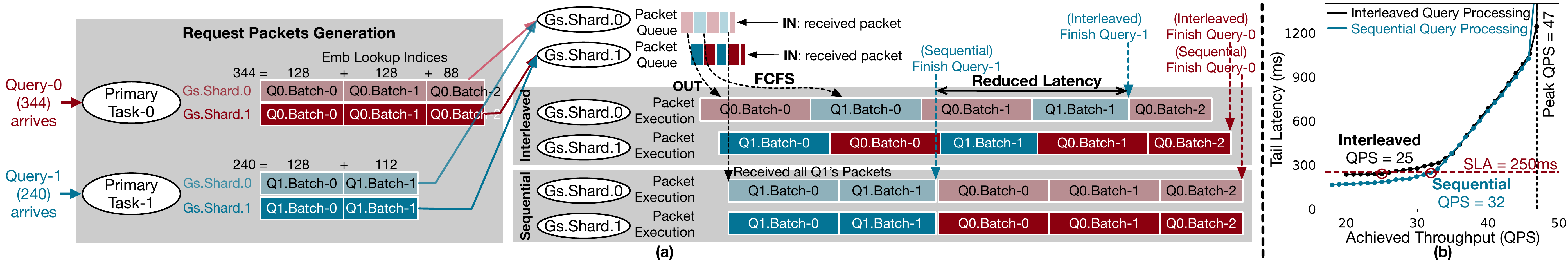}
  \vspace{-0.8cm}
  \caption{(a) Embedding operations' requests generation in primary tasks on two CNs; Interleaved and sequential query processing in SparseNet serving shards on two MNs;
  (b) Performance comparison of interleaved and sequential query processing.}
  \label{fig:async.sync}
  \vspace{-0.5cm}
\end{figure*}

\textbf{Managing Tasks.}
Adopting the distributed inference scheme, one serving unit consists of \{$n$ CNs, $m$ MNs\}.
There are $n$ primary inference tasks launched on the $n$ CNs, and $m$ remote SparseNet serving shards launched on the $m$ MNs.
Every primary task manages a private local memory region and a shared remote memory region. 
A CN's local memory and a GPU's HBM are mapped to the private local memory region.
The remote $M$ MNs' memory is mapped to the shared remote memory region.

%\textbf{Individual Task.}
Illustrated in Figure~\ref{fig:disagg.system}, a primary task with a unique global task ID (TID) is launched on every CN with dedicated CPUs for preprocessing $G_P$ and GPUs for DenseNet $G_D$. SparseNet RPC client operators are launched in the primary task. Based on the MemAccess routing table, the input data ({\em i.e.}, embedding lookup indices) of all embeddings are partitioned into $m$ packets and sent to $m$ designated MNs via RDMA writes. MNs respond to the primary CN with an acknowledgment signal and the remote memory addresses of the results (embeddings' Fsum). After receiving acknowledgments from all SparseNet shards, the primary CN loads results from remote MNs to the local GPU's HBM memory via RDMA reads. Then, DenseNet is launched on the GPU to calculate the final prediction.

\textbf{Handling Failures.}
CN and MN failures are independent and handled separately in a disaggregated serving unit (unlike failures in monolithic servers). 
In Figure~\ref{fig:process.memory.managment}(b), when a CN in a serving unit fails, only the primary task running on that CN is affected and migrated to a backup, over-provisioned node. The other tasks in the serving unit are unaffected.

When an MN fails, there are two possible scenarios. First, when there are multiple replicas of embedding tables allocated in the $m$ MNs, at least one copy of each embedding table is likely still available in the remaining MNs. We only need to update the MemAccess routing table, re-running the greedy routing to evenly distribute the embedding accesses on the remaining MNs. 
Second, when multiple MN failures lead to a loss of all replicas for an embedding table, memory is re-initialized to re-distribute all embedding tables across MNs after adding backup MNs to the serving unit.

\revise{
\textbf{Provisioning Heterogeneous Components.}
\note{add more specific techniques we do for heterogeneity}
Many emerging hardware technologies, such as compute-centric accelerators~\cite{CrossStackRec} and near-memory processing solutions~\cite{AxDIMM}, benefit recommendation performance.
However, deploying new components in existing datacenter infrastructure is a cumbersome process.
As datacenters deploy new servers and platforms to host specific devices, such as ZionEX to host GPUs~\cite{rec-training-ZionEX}, 
system architects are exposed to many produces from varied vendors.
%and need to identify the optimal combination of hardware components constructing the system before massive deployment.
Resource disaggregation simplifies how new hardware is deployed by introducing a new resource pool and allowing workloads to request desired hardware combinations from multiple pools. 
%Resource disaggregation simplifies a new hardware deployment as a new type of resource pool.
%Datacenter workloads can freely allocate the desired combination of compute and memory resources from these disaggregated pools.

%For example, the new compute-centric accelerator, e.g. the new generations of Nvidia GPUs, can be deployed as a new type of CN pool. Also, the novel NMP technology can be deployed as a new type of MN pool.
}

%\subsection{Intelligent Embedding Table Placement}
\subsection{Intelligent Embedding Management}

\note{
Talk about why a simple (random?) placement would not work, the challenges in placing embedding tables to multiple memory nodes: balance memory capacity utilization, memory bandwidth utilization, and reliability (replicas on different nodes)...}

To balance memory capacity, bandwidth utilization, and fault tolerance, we perform greedy embedding allocation and memory access routing during task initialization.
Embedding tables are read-only during inference, so remote memory does not cause data consistency and correctness issues.

\textbf{Embedding Allocation.} 
In Figure~\ref{fig:process.memory.managment}(a), one SparseNet shard contains a subset of embedding tables and $m$ shards are allocated to $m$ MNs. We take the embedding table as the basic unit for memory allocation and greedily assign embedding tables to the $m$ MNs following the scheme in Figure~\ref{fig:process.memory.managment}(c). 
Given the memory capacity provided by the $m$ MNs, the algorithm calculates the number of embedding replicas ($nReplicas$) that can be held by the $m$ MNs. Then, the algorithm picks the top $nReplicas$ MNs, ranked by available capacity, to allocate the $nReplicas$ of each embedding table.
The replicas of an embedding on different MNs provide the backup when memory node failure happens.

\textbf{MemAccess Routing.} 
All embeddings' memory accesses must be routed from the $n$ CNs to the $m$ MNs. We construct a MemAccess routing table that distributes memory accesses to the $m$ MNs. 
As depicted in Figure~\ref{fig:process.memory.managment}(c), for every embedding's memory access, the destination is picked from the $nReplicas$ MNs where that embedding table is allocated. 
First, the memory accesses of each individual embedding table is calculated by the average pooling factor multiply the embedding entry's dimension. The average pooling factor is profiled from embedding pooling operations shown in historical queries.
The greedy method selects the destination MN in the $nReplicas$ MNs that has the minimal memory accessed have been routed to. Once an MN has been selected, the tuple (task ID, embedding table ID, the destination MN) is added as an entry to the MemAccess routing table.

\revise{
\textbf{Why Not Random?} 
A naive method would randomly pick $nReplicas$ of MNs to allocate an embedding table, and also randomly pick the destination MN from the $nReplicas$ MNs to route the memory accesses.
In Figure~\ref{fig:process.memory.managment}(d), thousands of embedding tables are allocated on 8 MNs.
The random embedding management leads to unbalanced memory capacity allocation and memory accesses among the 8 MNs whereas our greedy method balances accesses among the 8 MNs.
}

\revise{
%\textbf{Global Management.}
\subsection{Interleaved versus Sequential Query Processing}
\note{Move the embedding table access scheduling: interleaved vs. sequential here (Figure 6) and move the comparison between the interleaved vs. sequential processing (Figure 5(a)) into evaluation.}

{\DesName} uses a task manager to coordinate scheduling within one serving unit. The manager can either perform interleaved or sequential query processing. In distributed inference, we find that sequential query processing offers lower latency and sustains higher throughput while satisfying the SLA. 

%the performance target (latency-bounded throughput) is heavily impacted by the scheduling scheme among distributed nodes. Sequential query processing offers lower latency and sustains higher throughput while satisfying the SLA. 

Figure~\ref{fig:async.sync}(a) illustrates one serving unit with \{2 CNs, 2 MNs\}. The embeddings' request packets are generated from primary tasks on the two CNs and forwarded to SparseNet shards on the two MNs.
\textit{Interleaved query processing} executes the packets on SparseNet shards in first-come-first-serve (FCFS) order, which seems like a natural design to maximize throughput. Under this scheme, packets from different queries are interleaved and both queries finish late. In contrast, \textit{sequential query processing} starts query execution only after all of its packets are received on all SparseNet shards. This scheme processes packets in lock step to finish one query's embedding operations together, allowing one of the queries to finish earlier. 

As shown in Figure~\ref{fig:async.sync}(b), interleaved and sequential query processing achieve similar peak throughput if ignoring the latency target. However, when the latency SLA at 250ms needs to be met, the latency-bounded throughput of the sequential scheme is 28\% higher than that of the interleaved scheme.

{\DesName}'s global manager performs sequential query processing to maximize latency-bounded throughput. In one serving unit, after all input data (embedding indices) belonging to one query are transferred from one CN to $m$ MNs, the global manager starts the query's embedding operations on the $m$ MNs simultaneously. After $m$ MNs finish all embedding operations for that query, the global manager schedules the $m$ MNs to proceed to the next query. The naive adoption of previous disaggregated designs~\cite{legoos,farview} usually considers interleaved query processing among remote MNs, allowing an MN to respond to multiple packets (for different queries) at the same time to maximize remote memory utilization. However, we find that such interleaving will harm response latency. 
}

%\subsection{Dynamic Resource Allocation at Scale}
\subsection{Failure-Aware Resource Allocation}
\label{sec:cluster_management}

\note{Move Figure 9 here. Emphasize the failure rate difference between GPU and CPU.}

To guarantee quality-of-service for billions of users, cluster resource management must allocate a sufficient number of serving units to ensure availability of the recommendation system. One serving unit consists of \{$n$ CNs, $m$ MNs\} in the disaggregated cluster, or $n$ servers in the monolithic server-based cluster, as one group to serve a recommendation model.
We formulate resource allocation as a constrained optimization problem with two main steps---offline workload characterization and online resource allocation.
The optimization's objective is the efficient allocation of serving units to achieve performance and availability goals. 

During offline workload characterization, we measure the throughput and power for each model-system pair, the achieved latency-bounded throughput $QPS_{M,S}$ and the peak power consumption $Power_{M,S}$ of model $M$ launched on one serving unit $S$, are input parameters when optimizing resource allocations for online serving.

During online resource allocation, subject to two constraints, Equation (1) allocates $N$ serving units (system type $S$ to model $M$) to minimize the total cost of ownership (TCO) for a time period t (e.g., 10s of minutes) where Capex is the acquisition cost of physical machines and Opex captures the operational cost due to electricity. Constraint (2) states the number of allocated serving units must be sufficient to guarantee availability given the highly-fluctuating diurnal loads and the expected machine failure rates. Backup machines are over-provisioned by $R$\% based on historical data on load variance and also by $F$\% given a machine failure rate. Constraint (3) states that provisioned power should be sufficient for allocated serving units.

\revise{
\vspace{-0.5cm}
\begin{align*} 
\text{Minimize}\ & N_{peak} Capex_S + \sum_{t}P(t) Rate_E & (1)\\
\text{subject to}\ & N(t) \geqslant (1+R\%) \frac{load(t)}{QPS_{M, S}} + & (2)\\
& \quad\quad\quad\ \frac{F_{CN}\% \cdot n + F_{MN}\% \cdot m}{n+m}  \cdot \frac{load_{peak}}{QPS_{M, S}} &\\
& P(t) \geqslant Power_{M, S} N(t) & (3)
\end{align*}
\vspace{-0.5cm}
}

\begin{figure}[t!]
  \centering
  \includegraphics[width=\columnwidth]{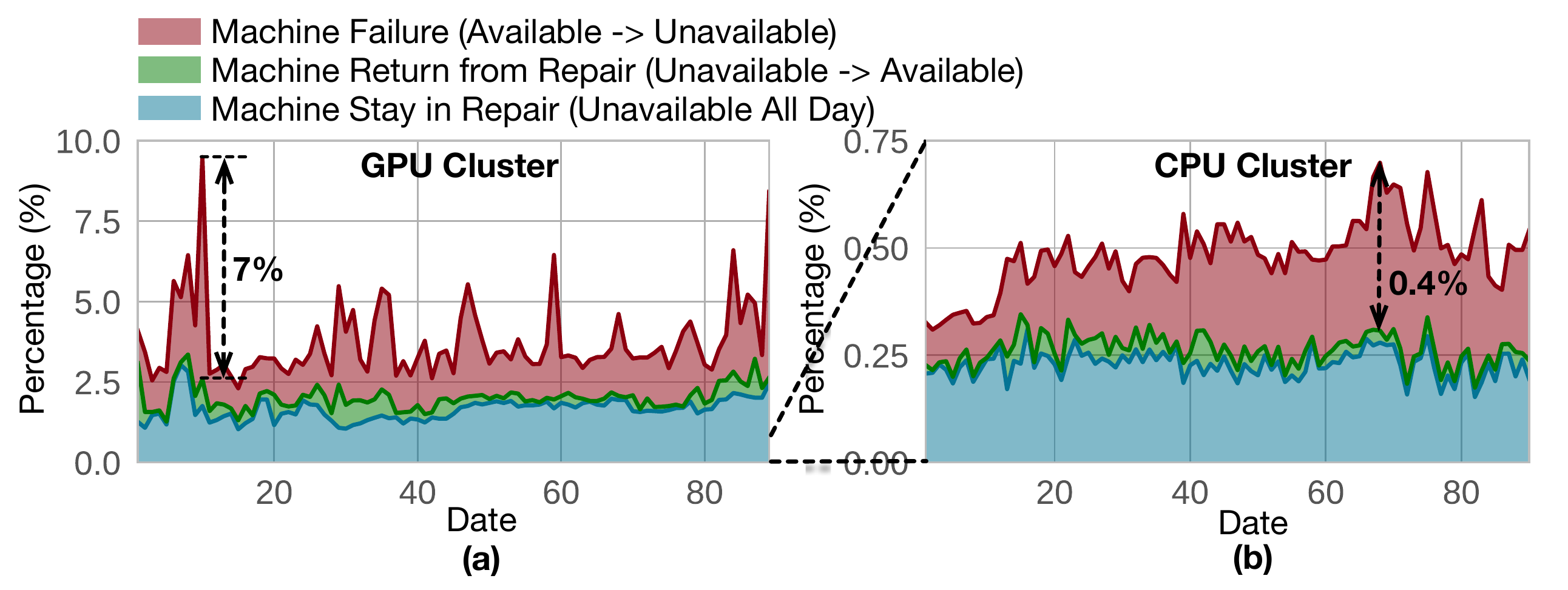}
  \vspace{-0.85cm}
  \caption{Daily machine failure rate of (a) GPU cluster and (b) CPU cluster.}
  \label{fig:server.failure}
  \vspace{-0.6cm}
\end{figure}

The failure rate for a machine is dictated by the failure rate of its least reliable component. 
When any component fails, the whole machine becomes inaccessible and its jobs must migrate to a backup server.
Figure~\ref{fig:server.failure} details reliability and failure rates in a production datacenter fleet. It reports four types of machine states during a day.
\begin{itemize}
\item Server available all day
\item Server inaccessible all day (\textcolor{blue}{blue} region)
\item Server becomes available mid-day (\textcolor{green}{green} region)
\item Server initially available but fails during day (\textcolor{red}{red} region)
\end{itemize}
We guarantee service availability by provisioning backup machines, which assume the responsibilities of failed servers in the fourth category. 

Monolithic servers handle failures inefficiently because they bundle resources together despite the distinct reliability characteristics of each system component. We find that CPU servers (with high-capacity memory systems) and GPU servers (with multiple powerful compute GPUs) exhibit very different failure rates. The daily failure rate for GPU servers (7\%) is much higher than that for CPU servers (0.4\%) in the historical, 90-day datacenter logs. 
The monolithic SU-2S/SO-1S server deploys CPUs, GPUs, and DRAMs together into single nodes, their failure rates follow the higher rate of 7\%. This higher failure rate implies more conservative over-provisioning, specifically on CPU and DRAM.

{\DesName} separates these unnecessarily bundled failures via a disaggregated architecture. It exploits the distinct failure rates of CNs and MNs to reduce the over-provisioning factor, in particular, for the more reliable MNs.

\if 0
\subsection{Support for Heterogeneity at Scale}
--> Overview
Many emerging hardware technologies have shown their performance improvement for recommendation workloads, including the compute-centric accelerator~\cite{CrossStackRec} and near-memory processing solutions~\cite{AxDIMM}.
However, deploying these new hardware devices in existing datacenter infrastructure is a painful process.
Datacenters always need to bring new servers as the platform to host specific hardware devices, such as ZionEX to host GPUs~\cite{rec-training-ZionEX}.
System architects are exposed to many hardware choices from different hardware vendors, and need to identify the optimal combination of hardware components constructing the system before massive deployment.

Resource disaggregation simplifies the new hardware deployment as a new type of resource pool in datacenters.
Datacenter workloads can freely allocate the desired combination of compute and memory resources from these disaggregated pools.
For example, the new compute-centric accelerator, e.g. the new generations of Nvidia GPUs, can be deployed as a new type of CN pool. Also, the novel NMP technology can be deployed as a new type of MN pool.
\fi

\section{Experimental Methodology}
\label{sec:method}

\begin{table*}[t!]
\scriptsize
\centering
\caption{System Parameters and Configurations}

\vspace{-0.25cm}
\begin{tabular}{|c|cccc|cccc|}
\hline
\multirow{2}{*}{} & \multicolumn{4}{c|}{Scale-Up and Scale-Out Monolithic Servers} & \multicolumn{4}{c|}{Disaggregated Nodes} \\ \cline{2-9} 
 & \multicolumn{1}{c|}{SU-2S} & \multicolumn{1}{c|}{\begin{tabular}[c]{@{}c@{}}SO-1S\\ Type I\end{tabular}} & \multicolumn{1}{c|}{\begin{tabular}[c]{@{}c@{}}SO-1S\\ Type II\end{tabular}} & \begin{tabular}[c]{@{}c@{}}SO-1S\\ Type III\end{tabular} & \multicolumn{1}{c|}{\begin{tabular}[c]{@{}c@{}}CN\\ Type I\end{tabular}} & \multicolumn{1}{c|}{\begin{tabular}[c]{@{}c@{}}CN\\ Type II\end{tabular}} & \multicolumn{1}{c|}{DDR-MN} & NMP-MN \\ \hline
\# Sockets & \multicolumn{1}{c|}{2$\times$IceLake} & \multicolumn{3}{c|}{1$\times$IceLake} & \multicolumn{2}{c|}{1$\times$CooperLake} & \multicolumn{2}{c|}{1$\times$ASIC Accelerator} \\ \hline
\# Channels per Socket & \multicolumn{1}{c|}{8} & \multicolumn{3}{c|}{8} & \multicolumn{2}{c|}{4} & \multicolumn{2}{c|}{8} \\ \hline
\# DIMMs per Channel & \multicolumn{1}{c|}{2} & \multicolumn{3}{c|}{2} & \multicolumn{2}{c|}{1} & \multicolumn{2}{c|}{2} \\ \hline
DRAM Device & \multicolumn{1}{c|}{\begin{tabular}[c]{@{}c@{}}DDR4-DIMM\\ 64GB, 3200MHz\end{tabular}} & \multicolumn{3}{c|}{\begin{tabular}[c]{@{}c@{}}DDR4-DIMM\\ 64GB, 3200MHz\end{tabular}} & \multicolumn{2}{c|}{\begin{tabular}[c]{@{}c@{}}DDR4-DIMM\\ 16GB, 2400MHz\end{tabular}} & \multicolumn{1}{c|}{\begin{tabular}[c]{@{}c@{}}DDR4-DIMM\\ 64GB, 3200MHz\end{tabular}} & \begin{tabular}[c]{@{}c@{}}NMP-DIMM\\ 64GB, 3200MHz\end{tabular} \\ \hline
GPU & \multicolumn{1}{c|}{8$\times$A100} & \multicolumn{1}{c|}{4$\times$A100} & \multicolumn{1}{c|}{2$\times$A100} & 1$\times$A100 & \multicolumn{1}{c|}{4$\times$A100} & \multicolumn{1}{c|}{1$\times$A100} & \multicolumn{2}{c|}{-} \\ \hline
NIC & \multicolumn{1}{c|}{2$\times$Front-End} & \multicolumn{3}{c|}{1$\times$Front-End, 1$\times$Back-End} & \multicolumn{2}{c|}{1$\times$Front-End, 1$\times$Back-End} & \multicolumn{2}{c|}{1$\times$Back-End} \\ \hline
\end{tabular}

\label{tab:sys_config}
\vspace{-0.5cm}
\end{table*}

\begin{table}[t!]
\scriptsize
\centering
\caption{Commodity Hardware Devices}

\vspace{-0.3cm}
\begin{tabular}{|c|c|c|c|}
\hline
 Devices & Specs & TDP & Market Price \\ \hline
IceLake CPU & \begin{tabular}[c]{@{}c@{}}40 Cores @ 2.30GHz\\ Intel Xeon Platinum 8380\end{tabular} & 270 W & \$4K$\sim$\$5K \\ \hline
CooperLake CPU & \begin{tabular}[c]{@{}c@{}}26 Cores @ 1.40GHz\\ Intel Xeon Platinum 8321HC\end{tabular} & 86 W & \$2K$\sim$\$3K \\ \hline
GPU & Nvidia A100 (80GB HBM) & 400 W & \$12K$\sim$\$15K \\ \hline
\multirow{2}{*}{DDR4-DIMM} & 16GB, 1 Rank @ 2400MHz & 5 W & $\sim$\$80 \\ \cline{2-4} 
 & 64GB, 2 Ranks @ 3200MHz & 24 W & $\sim$\$350 \\ \hline
NMP-DIMM & 64GB, 2 Ranks @ 3200MHz & 24 W & Assume $\sim$\$700 \\ \hline
NIC & \begin{tabular}[c]{@{}c@{}}Mellanox ConnectX-6\\ @ 200Gbps\end{tabular} & 20 W & $\sim$\$2.5K \\ \hline
\end{tabular}

\label{tab:devices}
\vspace{-0.5cm}
\end{table}

\subsection{Measuring Monolithic Servers}

Table~\ref{tab:sys_config} describes GPU servers used to evaluate monolithic servers for recommendation systems. These servers are used to evaluate model inference on the state-of-the-art (SOTA) baseline, (a) end-to-end inference on a single scale-up server, and (b) distributed inference on multiple scale-out servers. 

\textbf{SU-2S} is a scale-up dual-socket server with two 40-core Intel IceLake CPU processors, two terabytes (TB) of memory, and eight NVIDIA A100 GPUs, each with 80GB HBM. Front-end NICs are attached to the two CPU processors to receive inference queries from the front-end networks. 

\textbf{SO-1S} is a scale-out single-socket server with one 40-core Intel IceLake CPU processor, one terabyte (TB) of memory, and four Nvidia A100 GPUs. One front-end NIC is attached to the CPU processor to receive inference queries from the front-end networks. Two back-end NICs are connected through the PCIe switch to communicate intermediate model inference data, embedding indices and Fsum, between the servers multiple SO-1S servers. We emulate different types of SO-1S servers by limiting GPU usage to one, two, or four GPUs.

\subsection{Emulating {\DesName} System} 

Table~\ref{tab:sys_config} describes two types of compute nodes (CNs) and memory nodes (MNs) that we consider for the disaggregated system design. We emulate the recommendation system on disaggregated CNs and MuccjbhdcgblnjgulkklltdkkbbjguevbjlkjnjdbnjreedvfcvcdevigiNs by launching stages of the inference pipeline on the CPU and GPU servers. All performance numbers are measured from real systems except those with NMP-enabled memory, which is estimated with cycle-level simulation. 

\textbf{Compute Nodes.} One CN is configured with one light-weight CPU and either one or four GPUs, matching the number of GPUs in the two SO-1S configurations. Each CN has one front-end NIC that receives incoming inference queries. It also has one back-end NIC that sends embedding indices to MNs and loads the embedding Fsum from MNs.

CN emulation accounts for two stages of the inference pipeline. Preprocessing $G_P$ is launched on the light-weight CPU server with one Intel CooperLake CPU and 64GB memory. DenseNet $G_D$ is launched on an SO-1S server using the selected number of GPUs.

\textbf{Memory Nodes.} First, the DDR-MN is configured with 1TB memory, matching the memory capacity of the SO-1S server, and a light-weight ASIC as the MN-side processing unit. Alternatively, the NMP-MN represents a memory node with near-memory processing enabled (NMP-DIMMs)~\cite{RecNMP,AxDIMM}. The NMP-MN exploits DIMM- and rank-level parallelism to increase effective memory bandwidth by $4\times$ relative to DDR-MN. DIMM-level parallelism doubles bandwidth by provisioning two NMP-DIMMs on a single memory channel. Rank-level parallelism further doubles the bandwidth by provisioning two ranks on one NMP-DIMM. We conservatively estimate the power dissipated by the MN's ASIC to be 23.9 Watts based on the power profile of an internal ASIC accelerator fabricated in TSMC 7nm. 

DDR-MN emulation launches SparseNet $G_S$ on SO-1S while disabling all GPUs. NMP-MN emulation simulates emerging NMP-DIMMs, following the earlier methodology for NMP studies~\cite{tensordimm,RecNMP,tensorcasting,FAFNIR}.

\textbf{Communication.} High-performance communication between CNs and MNs is the key for efficiency. One serving unit is defined by $n$ CNs and $m$ MNs and we explore a range of design points, ranging from the minimum (1 CN, 1 MN) to the maximum (8 CNs, 8 MNs). We emulate communication within the recommendation serving unit by microbenchmarking communication between 16 (8 + 8) GPUs on two SU-2S servers. Communication uses NVIDIA’s Collective Communication Library (NCCL)~\cite{nccl} and RDMAs. For each inference task, a CN sends embedding indices to multiple MNs with a \textit{Scatter} (one-to-all) operation and the CN loads embedding Fsums from multiple MNs with a \textit{Gather} (all-to-one) operation.

\subsection{Total Cost of Ownership}

We estimate the total cost of ownership (TCO) using public market prices for all commodity devices and assuming a three-year machine lifetime. 
%Since there is no NMP-DIMM readily available in our datacenters, 
\revise{Since there is no NMP-DIMM commercially available on the market,}
we assume the price of one NMP-DIMM is 2$\times$ that of a regular DDR-DIMM because NMP-DIMM can theoretically double effective memory bandwidth to the same memory capacity. This estimate is conservative because NMP-DIMM will be a value-added product over today's DIMM with only a small, fractional increase in cost when the technology is standardized. %% by JEDEC.

\subsection{Evaluation Framework} 

For monolithic servers, the performance of model inference on the scale-up single server and the scale-out multiple servers are evaluated on real systems, building atop of the DeepRecSys serving framework~\cite{DeepRecSys}.

For disaggregated systems, latency and power dissipated in the three pipeline stages are recorded for each inference query. Latency in the three stages---preprocessing $G_P$ on a CPU, SparseNet $G_S$ on DDR-MN, and DenseNet $G_D$ on GPUs---are individually measured and recorded on real systems. During real model inference execution, the dummy serving processes of the corresponding pipeline stages are replayed with the recorded latency for each query for each pipeline stage. When NMP-MNs are evaluated, cycle-level simulation estimates SparseNet $G_S$ performance. CPU and DDR4 power is read from Intel RAPL~\cite{RAPL}, and GPU power is measured by Nvidia API \textit{nvidia-smi}.
\section{Evaluation Results}
\label{sec:eval}

%Goal: explain how we find the best design point
%More detailed points to make: 
% 1) Why more memory notes per serving unit is better - lower embedding acccess latency.
% 2) Reliability difference between compute and memory nodes in (d)

% NMP - 2 DIMMs per channel. For each DIMM, two ranks activated at the same time. This leads to 2x2 = 4 times higher effective bandwidth.

\revise{
\subsection{Substantial Cost Increase with Model Evolution}
\label{sec:hetero}

%\textit{Varying compute and memory demands of diverse recommendation models fundamentally defy a "one-size-fit-all" system solution. System heterogeneity permits server configurations that are optimized for memory-intensive RM1 and compute-intensive RM2.}

\textit{Takeaway$_A$: While system heterogeneity permits server configurations to be tailored for diverse recommendation models, the exponential model growth drives up the datacenter costs substantially.}
%Thus, system heterogeneity needs to be offered to minimize the cost that different server configurations are optimal for the memory-dominated RM1 and compute-dominated RM2. 
%thus a high degree of system heterogeneity needs to be offered in monolithic server-based clusters to meet such demands and reduce the cost.

We consider recommendation services RM1 and RM2 and their respective model evolution for the next three years, as illustrated in Figure~\ref{fig:model.projection}(c).
We explore the server design space and optimize monolithic server configurations to minimize the TCO subject to performance goals. 
We consider five system configurations, including naive and NUMA-aware model inference on a SU-2S server as well as distributed inference on three types of SO-1S servers with 1, 2, and 4 GPUs. We follow method in Sec.~\ref{sec:cluster_management} to estimate the cluster TCO for online serving based on offline measured throughput and power consumption. %TCO accounts for the number of servers required (Capex) and their power dissipation (Opex). 

\begin{figure}[t!]
  \vspace{-0.1cm}
  \centering
  \includegraphics[width=\columnwidth]{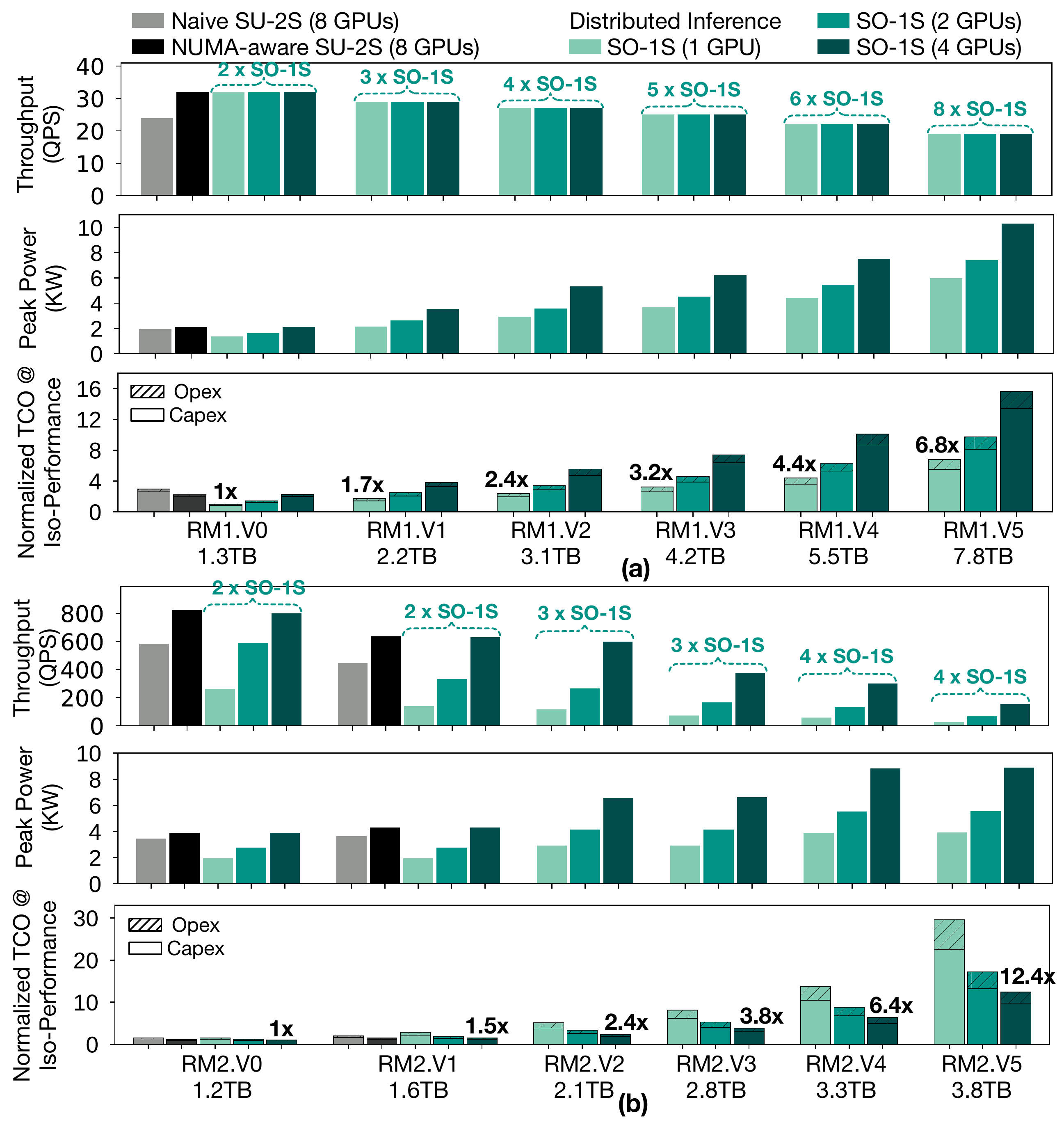}
  \vspace{-0.85cm}
  \caption{Latency-bounded throughput (QPS), measured power consumption (KiloWatt), and normalized TCO serving the same incoming loads of (a) RM1.V0 -- V5 and (b) RM2.V0 -- V5 on a SU-2S server and a group of distributed SO-1S servers deployed 1, 2, 4 GPUs.}
  \label{fig:model.evolution}
  \vspace{-0.4cm}
\end{figure}

\begin{figure}[t!]
  \centering
  \includegraphics[width=\columnwidth]{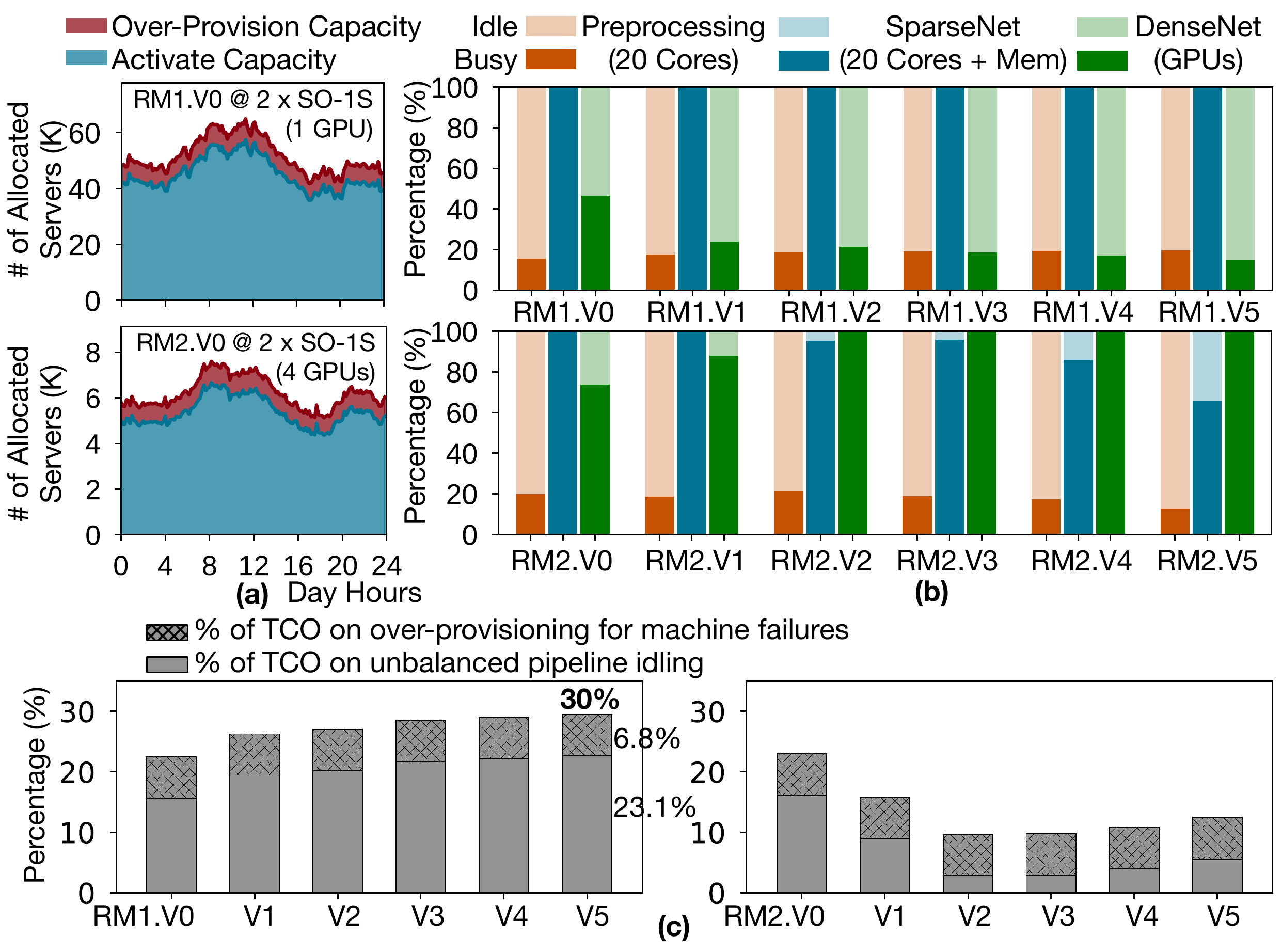}
  \vspace{-0.85cm}
  \caption{(a) Daily cluster capacity provisioning;
  (b) Resource under-utilization from unbalanced pipeline idleness;
  (c) The percentage of total cluster TCO cost on the over-provisioned capacity and unbalanced pipeline idleness.}
  \label{fig:failure.idleness}
  \vspace{-0.6cm}
\end{figure}

In Figure~\ref{fig:model.evolution}, when the model size grows larger than 2TB, one SU-2S server no longer provides sufficient memory capacity, and distributed inference is required. The number of distributed SO-1S servers in one serving unit is determined by the model size.
For memory-intensive RM1, distributed inference on SO-1S (1 GPU) servers is the most cost-effective. For compute-intensive RM2, distributed inference on SO-1S (4 GPUs) servers is the best system choice. 
Thus, a heterogeneous cluster based on monolithic design would host two types of servers.
Despite deploying these heterogeneous servers optimize for each recommendation service, as models grow in size and complexity, the datacenter's TCO increases substantially---by 6.8$\times$ for RM1 and 12.4$\times$ for RM2---over the three-year period.

\subsection{Cost Inefficiency from Monolithic Design }
\label{sec:inefficiency}

\textit{Takeaway$_B$: Because monolithic servers bundle compute and memory in a fixed ratio, up to 30\% of datacenter cost is wasted on idle resources (23.1\%) and over-provisioned capacity (6.8\%). These inefficiencies arise due to unbalanced model pipelines and heterogeneous fault rates.}
}

Figure~\ref{fig:failure.idleness}(a) shows server allocation and utilization in the span of a day. The blue region shows the activated capacity for serving queries while the red region represents the over-provisioned capacity assuming a 7\% machine failure rate observed from a production datacenter fleet. We assume servers are abundant and workloads always receive their preferred machines, optimally SO-1S (1 GPU) for RM1 and SO-1S (4 GPUs) for RM2.

\revise{Figure~\ref{fig:failure.idleness}(b) shows resource utilization for active servers.
RM1 models are constrained by SparseNet (dark blue for busy) while CPUs and GPUs are under-utilized during the Preprocessing and DenseNet stages (light orange and light green for idling), respectively.
RM2 models are first constrained by SparseNet (dark blue), and then by DenseNet (dark green), as RM2 models grow primarily in the DenseNet.}

Figure~\ref{fig:failure.idleness}(c) shows the percentage of TCO wasted on idle resources attributed by two sources. We assume the CPU costs for carrying out Preprocessing and SparseNet are the same. For both models, over-provisioned capacity accounts for 6.8\% of TCO. In addition, RM1's unbalanced pipeline costs 15.6\%--23.1\% of TCO while RM2's costs 2.8\%--16.2\%. RM1 pays higher for idleness because its DenseNet computation poorly utilizes the expensive GPUs; in contrast, RM2 only pays the toll in less expensive CPUs and memories during Preprocessing and SparseNet.

\begin{figure*}[t!]
  \centering
  \includegraphics[width=\textwidth]{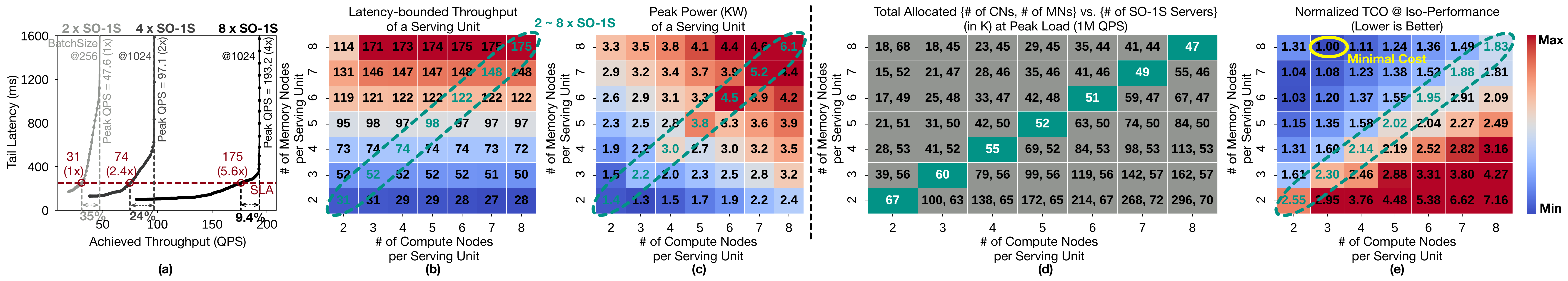}
  \vspace{-0.8cm}
  \caption{(a) Throughput-latency tradeoff of RM1.V0 launched on distributed 2, 4, 8 SO-1S servers (1 GPU);
  Task scheduling space exploration of RM1.V0 on distributed SO-1S servers and disaggregated \{CNs, DDR-MNs\}, including (b) latency-bounded throughput and (c) power consumption of one serving unit;
  Cluster capacity allocation: (d) total number of allocated SO-1S servers and \{CNs, DDR-MNs\} and (e) normalized cluster TCO to serve peak load.
  }
  \label{fig:task.schedule.space}
  \vspace{-0.6cm}
\end{figure*}

\subsection{Improving Cost Efficiency by Scaling Out}
%\note{Inefficiency \% and scale-out optimization in the monolithic server case}

%\revise{We explore distributed inferences for the inference serving unit of a monolithic server fleet and the proposed {\DesName}. 
%%on monolithic servers' and {\DesName}'s inference serving unit which is a collection of $n$ SO-1S servers and \{$n$ CNs, $m$ MNs\}.}

\textit{Takeaway$_C$: Scaling the number of monolithic servers used for distributed inference improves latency-bounded throughput and cost efficiency.}

\revise{
We explore the benefits of distributed inference and scaling out monolithic servers by examining the diagonals in Figure~\ref{fig:task.schedule.space}. A serving unit may scale out with two to eight monolithic SO-1S servers for RM1.V0. Adding more servers helps reduce query response latency, and then improves latency-bounded throughput for serving units that perform sequential (and not interleaved) query processing. 
In Figure~\ref{fig:task.schedule.space}(a), a serving unit with two, four, and eight SO-1S servers achieves 65\%, 76\%, and 90.6\% of that serving unit's peak throughput, respectively. Scaling out causes performance to increase superlinearly as throughput improves by 2.4$\times$ and 5.6$\times$ with 2$\times$ and 4$\times$ the number of servers. 

A serving unit's throughput and the over-provisioning ratio based on machine failure rates---which we take to be 7\% for SO-1S, 7\% for CNs, and 0.04\% for MNs---determine how many serving units are required to ensure the cluster satisfies peak query load (Figure~\ref{fig:task.schedule.space}(d)). 
The recommendation serving cluster's cost depends on the number of hosted serving units and their power costs (Figure~\ref{fig:task.schedule.space}(c)). Our analysis indicates that scaling out with monolithic serves causes normalized TCO to decrease from 2.55$\times$ to 1.83$\times$ (Figure~\ref{fig:task.schedule.space}(e), diagonal). 
}

%Figures~\ref{fig:task.schedule.space}(b), (c), and (e) present grids of throughput, power, and cost measurements for various serving unit configurations. The grid's diagonal represents configurations that consists of two to eight monolithic SO-1S servers, each providing a fixed bundle of compute and memory capacity. 

%Figure~\ref{fig:task.schedule.space}(d)(e) uses a serving unit's throughput to determine how many serving units must be allocated to ensure the cluster satisfies peak query load. Note that the number of allocated nodes includes over-provisioned capacity to ensure availability given machine failure rates, we take 7\% for SO-1S, 7\% for CNs, 0.04\% for MNs.

%The grid's diagonal consists of configurations that equally balance compute and memory using two to eight monolithic SO-1S servers for distributed inference. We vary the \{CN, MN\} configuration and optimize task schedules to determine the most efficient disaggregated system for serving recommendation models. 

%The cluster's \{CNs, MNs\} allocation dictates the recommendation system's total power and cost. TCO accounts for the number of nodes required (Capex) and their power dissipation (Opex). 

\subsection{Improving Cost Efficiency by Disaggregation}

\begin{figure}[t!]
  \vspace{-0.15cm}
  \centering
  \includegraphics[width=\columnwidth]{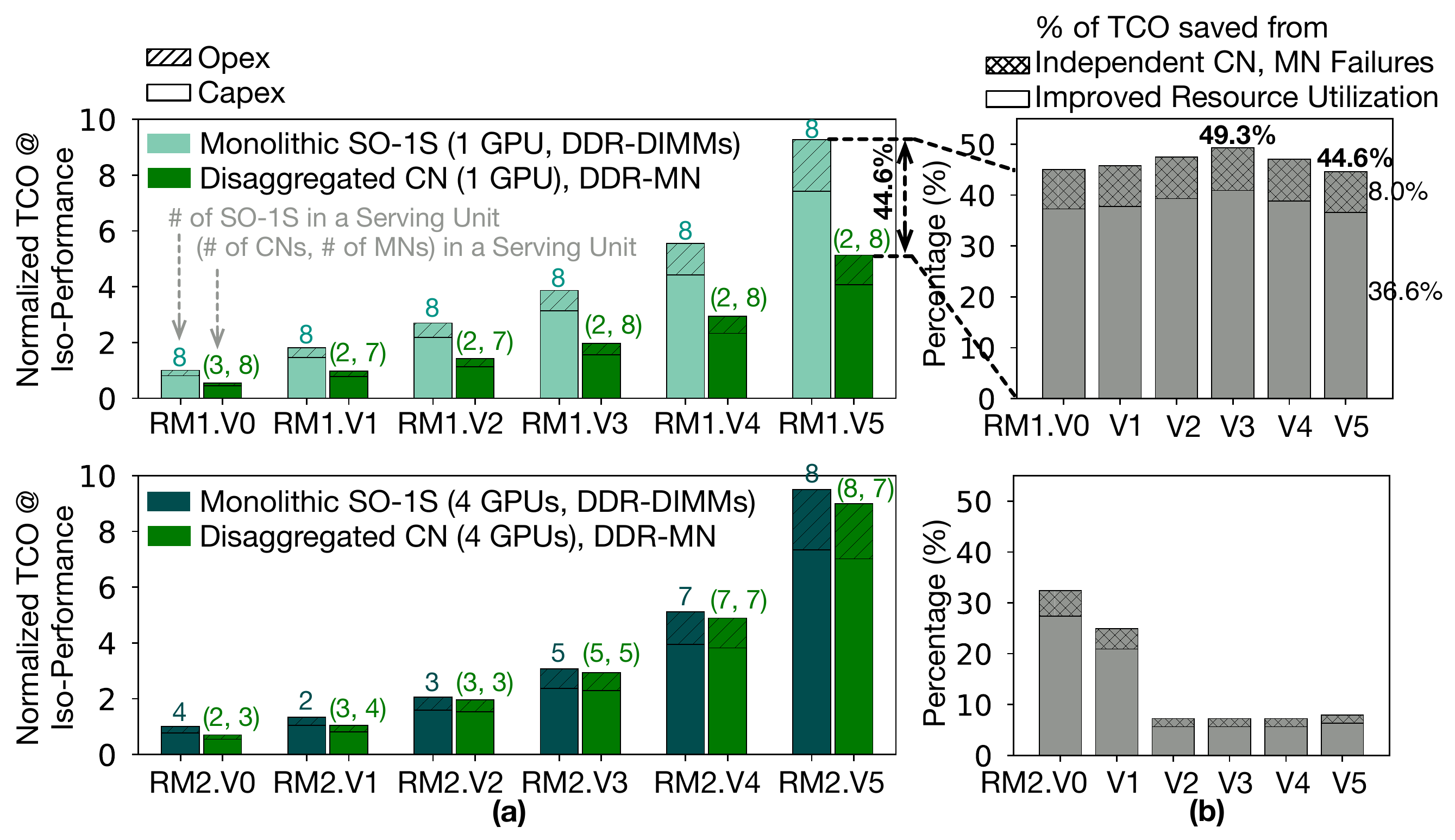}
  \vspace{-0.88cm}
  \caption{(a) Normalized cluster TCO of RM1.V0--V5 and RM1.V0--V5 launched on the distributed SO-1S servers, disaggregated CNs and DDR-MNs;
  (b) The percentage breakdown of saved TCO from improved resource utilization and independent compute and memory node failures.
  }
  \label{fig:tco.saving.disaggregation}
  \vspace{-0.6cm}
\end{figure}

\textit{Takeaway$_D$: Disaggregation further improves cost efficiency by improving resource utilization and reducing over-provisioned backups for machine failures.} 

We explore the benefits of distributed inference on disaggregated compute and memory nodes by examining the whole 2D grid in Figure~\ref{fig:task.schedule.space}. 
Disaggregation permits a serving unit to scale compute and memory nodes independently, which better matches resources to workload needs. 
For memory-intensive model RM1.V0, throughput is relatively insensitive to the number of CNs in the serving unit. Disaggregation permits a serving unit with fewer CNs and more MNs, which in turn reduces cost (Figure~\ref{fig:task.schedule.space}(e), blueish region). 
A serving unit that deploys 3 CNs and 8 MNs minimizes the TCO with negligible impact on performance. The throughput of this cost-efficient solution is only 2\% less than that of eight monolithic SO-1S servers (Figure~\ref{fig:task.schedule.space}(b)).

%\revise{Let us turn attention to the whole 2D grid in Figure~\ref{fig:task.schedule.space} representing the varying number of disaggregated CNs and MNs in a serving unit.
%The model inference on a disaggregated serving unit is also supported by distributed inference.}

%Moreover, it compares efficiency when scaling a system with monolithic servers, and disaggregated CNs and MNs. 

Broadening our evaluation, Figure~\ref{fig:tco.saving.disaggregation} details disaggregation's benefits for six generations of recommendation models. For memory-intensive RM1, disaggregation reduces the TCO by up to 49.3\% compared to distributed inference on monolithic SO-1S servers. Most of the saving (40.9\% of 49.3\%) comes from reducing the number of CNs, which are equipped with expensive GPUs. The remaining saving comes from exploiting MNs' low failure rates and over-provisioning resources by a smaller factor. 

For compute-intensive RM2, serving units require similar amounts of hardware whether using monolithic servers or using disaggregated CNs and MNs. For example, the optimal configurations for V2, V3, and V4 models use the same ratio of CNs and DDR-MNs. Yet even for these models, disaggregation reduces cost by 4.3\% to 9.3\%. 

The cost savings come from two sources. First, the serving unit with disaggregated nodes is more power-efficient. CNs deploy low-end CPUs and high-end GPUs. MNs deploy ASICs for a modest amount of computation near the data. These CN and MN configurations dissipate less power than the monolithic SO-1S servers deploying a high-end CPU (\textit{i.e.}, 40-core IceLake). Lower operating expenses account for 7.2\% of the cost savings. Second, the MN's lower failure rate permits the serving unit to over-provision resources by a smaller factor.

\revise{
\subsection{Provisioning Heterogeneity by Disaggregation}

\begin{figure}[t!]
  \centering
  \includegraphics[width=\columnwidth]{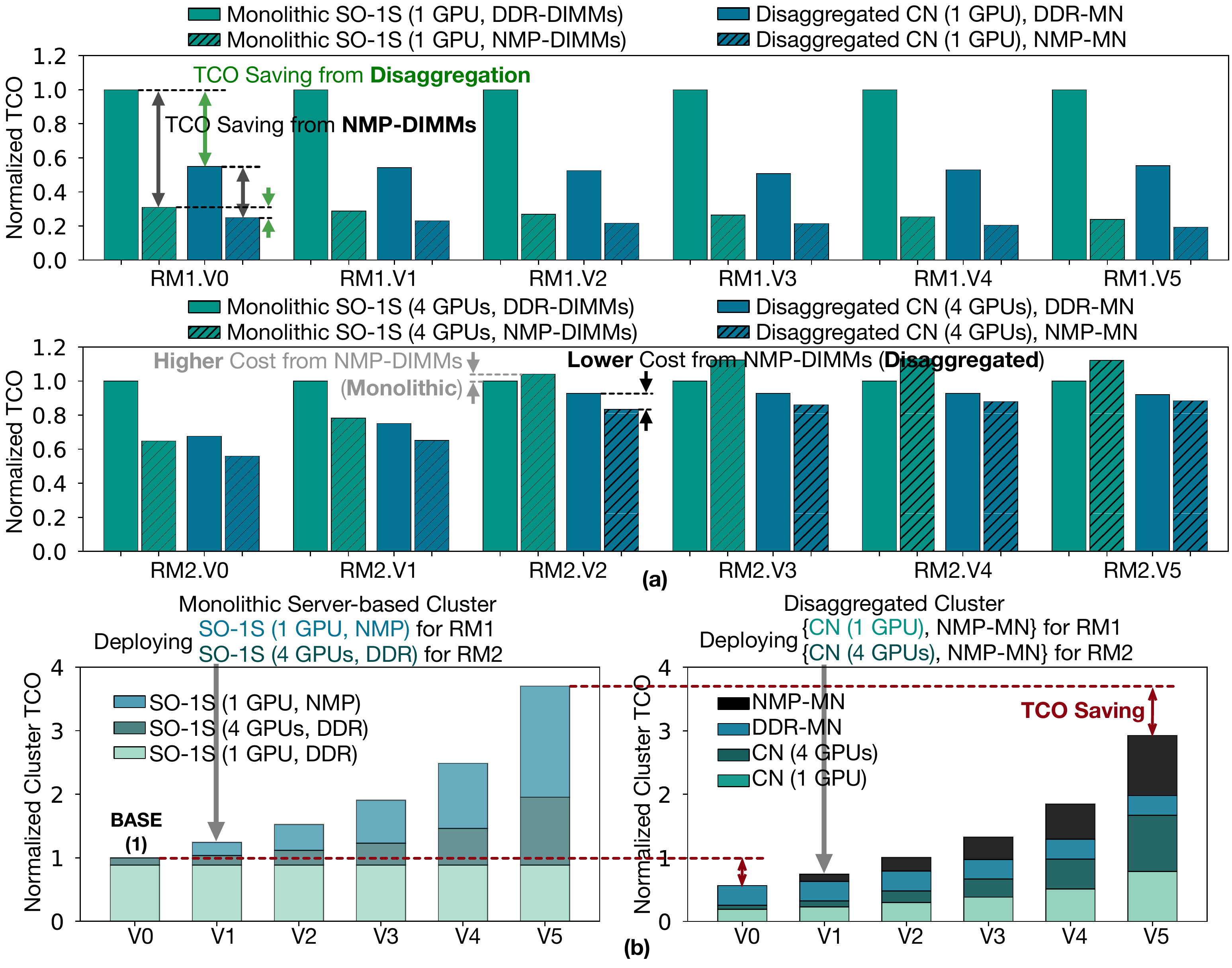}
  \vspace{-0.85cm}
  \caption{
  (a) System efficiency explorations for the 6 generations of RM1 and RM2 models.
  (b) Comparison of monolithic server-based cluster and disaggregated cluster along the three-year model evolution.
  }
  \label{fig:tco.saving.nmp}
  \vspace{-0.6cm}
\end{figure}

\textit{Takeaway$_E$: Resource disaggregation provides flexible support for resource heterogeneity in production datacenters. Hardware components are organized into disparate resource pools rather than integrated into monolithic servers. This organization improves utilization and reduces costs.}

% (1) Assumption: the initial cluster is built to optimally serve both RM1 and RM2 with the identified optimal choices of server configuration, XXX and XXX, for RM1 and RM2 respectively
% One important assumption is the initial deployed servers will continue exist for the future generations with 3-year machine lifetime.
% Moving forward, we introduce NMP as new type of device to expand the initial cluster. for monolithic cluster, this leads to two new server configurations, XX and YY.
% For disggregated cluster, it's a new type of MN, NMP-MN.

% in the similar manner as before, we evaluate the normalized TCO across the model generations with different system configurations, including XX and YY in 

%, hardware deployment can be simplified as a new type of resource pool, instead of the tightly bundled specific hardware devices on monolithic servers.

We explore technology scenarios, comparing and contrasting a cluster with monolithic servers and one with disaggregated resources as recommendation models evolve and grow over three years. 
The initial clusters are built with commodity CPUs, regular DDR-DIMMs, and GPUs.
%Each cluster's configuration is optimized to serve RM1 and RM2 models. 
%To optimally serve RM1.V0 and RM2.V0 models, the identified optimal system configurations are deployed for both clusters.
The cluster with monolithic servers optimally deploys SO-1S servers with 1 GPU and DDR-DIMMs for memory-intensive RM1.V0 and optimally deploys SO-1S servers with 4 GPUs and DDR-DIMMS for compute-intensive RM2.V0. 
%As such, for the monolithic server-based cluster, SO-1S servers with 1 GPU and DDR-DIMMs are deployed for memory-dominated RM1.V0, and SO-1S servers with 4 GPUs and DDR-DIMMs are deployed for compute-dominated RM2.V0.
On the other hand, the cluster with disaggregated compute and memory nodes optimally deploys \{CN with 1 GPU, DDR-MN\} and \{CN with 4 GPUs, DDR-MN\} for RM1.V0 and RM2.V0, respectively.
We assume that deployed servers and nodes will remain deployed for their three-year machine lifetimes.

%In the monolithic clusters, NMP-DIMMs are deployed with SO-1S server and bundled with 1 GPU / 4 GPUs, while in the disaggregated cluster, NMP-DIMMs are independently deployed as a new type of memory node or NMP-MN in Table~\ref{tab:sys_config}.

In future model generations, near-memory processing leads to new system components (NMP-DIMM). 
The cluster with monolithic servers deploys two new server configurations, namely an SO-1S server with 1 GPU and NMP-DIMMs as well as an SO-1S server with 4 GPUs and NMP-DIMMs.
On the other hand, the disaggregated cluster deploys NMP-DIMMs as a new type of memory node, NMP-MN (Table~\ref{tab:sys_config}).

%Moving forward for future model generations, RM1.V1$\sim$V5 and RM2.V1$\sim$V5, we consider the near-memory processing DRAM DIMM (NMP-DIMM) to be a new memory device type for enhancing and constructing the systems.
%Hence, for the monolithic server-based cluster, two new server configurations are deployed: SO-1S server with 1 GPU and NMP-DIMMs, and SO-1S server with 4 GPUs and NMP-DIMMs.
%For the disaggregated cluster, NMP-DIMMs are deployed as a new type of memory node, NMP-MN (Table~\ref{tab:sys_config}).

Figure~\ref{fig:tco.saving.nmp}(a) shows the TCO savings from NMP-DIMMs. %the provision of servers and nodes.
When monolithic servers are used, NMP-DIMMs reduce costs for memory-intensive RM1 models but increase costs for compute-intensive RM2 models. Emerging NMP-DIMMs are more expensive than conventional DDR-DIMMs starting in the V2 generation. Their costs are justified only for memory-intensive workloads that experience throughput gains when adopting the technology. 

Specifically, NMP-DIMMs increase effective memory bandwidth by 4$\times$ through DIMM- and rank-level parallelism. Greater memory bandwidth accelerates embedding operations.
The SO-1S server with NMP-DIMMs improves RM1 throughput by up to 3.64$\times$.
However, for compute-dominated RM2, the memory bandwidth of NMP-DIMMs on the SO-1S server is under-utilized, and the 2$\times$ cost of NMP-DIMMs (Table~\ref{tab:devices}) over DDR-DIMMs eventually leads to a higher TCO. 

In contrast, when disaggregated nodes are used, NMP-DIMMs reduce costs for both RM1 and RM2 models because the emerging technology (NMP-DIMM) is deployed as a new resource pool and allocated flexibly. 
The independent scaling of CNs and MNs in the disaggregated cluster allows a smaller ratio of MNs in a serving unit,
which prevents the under-utilization of NMP-DIMMs' memory bandwidth.

%The model-system efficiency exploration, shown in Figure~\ref{fig:tco.saving.nmp}(a), is performed to identify the optimal cluster provisioning.
%In the monolithic server-based cluster, NMP-DIMMs help reduce the cost for all memory-dominated RM1 models, whereas for compute-dominated RM2 models, NMP-DIMMs become costlier than regular DDR-DIMMs starting from V2 generation.
%In contrast, in the disaggregated cluster, deploying NMP-DIMMs achieves lower TCO for both RM1 and RM2 across the board.

%%Considering the fast-evolving recommendation models and the 3-year machine lifespan in datacenters, in 
%As indicated by Figure~\ref{fig:tco.saving.nmp}(b), the monolithic cluster expands the capacity by deploying the SO-1S (1 GPU, NMP-DIMMs) servers for RM1 and SO-1S (4 GPUs, DDR-DIMMs) for RM2 across model generations V1 to V5, while the disaggregated cluster expands the capacity by \{CN (1 GPU), NMP-MN\} for RM1 and \{CN (4 GPUs), NMP-MN\} for RM2.
%Overall, the disaggregated cluster achieves 27.2\%--43.4\% cost savings over the monolithic server-based cluster.

Figure~\ref{fig:tco.saving.nmp}(b) shows total cluster TCOs over multiple model generations when the cluster is provisioned by continuously deploying optimal system configurations for new generations of RM1 and RM2.
%including the SO-1S server with (1 GPU, NMP-DIMMs) for RM1 and the SO-1S server (4 GPUs, DDR-DIMMs) for RM2 in the monolithic server-based cluster, and \{CN (1 GPU), NMP-MN\} for RM1 and \{CN (4 GPUs), NMP-MN\} for RM2 in the disaggregated cluster.
When monolithic servers are used, the cluster expands capacity for evolving recommendation models by deploying SO-1S servers with 1 GPU and NMP-DIMMs for RM1.V1--V5, and deploying SO-1S servers with 4 GPUs and DDR-DIMMs for RM2.V1--V5.
When disaggregated resource nodes are used, the cluster deploys \{CN with 1 GPU, NMP-MN\} and \{CN with 4 GPUs, NMP-MN\} for RM1 and RM2, respectively.
%Overall, disaggregation reduces TCO by 21\%--43.6\% over the use of monolithic servers as models evolve over three years. 
Overall, the disaggregated cluster allows 21\%--43.6\% TCO saving over the monolithic server-based cluster across the three-year model evolution. % as models evolve over the three-year server lifetime.
%across the three-year model evolution.

%\note{Liu: Feel wired to explain the saved TCO value in text here, it's not absolute (more like absolute normalized TCO), nor percentage. Any better way?}
}

\section{Related Work}
\label{sec:relatedworks}

Disaggregating organizes different types of resources into separate pools for independent, fine-grained resource allocation. Several studies disaggregate datacenter storage, {\em e.g.}, ~\cite{alibaba-pangu,amazon-s3,facebook-bc,snowflake} where network communication can be hidden by the high storage latency. Given the success of disaggregated storage and rapidly evolving network technologies~\cite{ccix,genz,intel-omni,nvidia-InfiniBand,network-osdi-2016,lite,shoal,mind}, 
disaggregated memory systems were proposed for large-scale datacenters ~\cite{hp-themachine,hp-memory-driven-computing,firebox,dredbox,disaggmem-isca-2009,disaggmem-hpca-2012,legoos,disaggmem-eurosys-2018,minsoo-micro-2018,minsoo-IEEEMicro-2019,disaggmem-atc-2020,disaggmem-atc-2021,disaggmem-asplos-2021,farview,clio,farmem-nsdi-2014,farmem-eurosys-2020,farmem-osdi-2020}.
Disaggregated memory promises large memory capacity, independent scale-out for compute and memory resources, improved reliability by separating compute nodes' failures from memory nodes', more cost-efficient hardware deployment, etc. Disaggregated memory systems are particularly attractive for deep learning workloads as rapid growth in datasets and models turn DRAM into a major system bottleneck. 

Disaggregated systems may deploy memory nodes with and without processing capabilities. MNs without processing are viewed primarily as raw physical memory. Such MNs have been adopted in HPE’s Memory-Driven Computing project~\cite{hp-themachine,hp-memory-driven-computing}, the disaggregated hashing system~\cite{disaggmem-atc-2020} and disaggregated key-value systems~\cite{disaggmem-atc-2021}, \revise{also the recent compute express link (CXL)-based memory pooling system~\cite{Pond}. However, for memory-intensive workloads, data movement through the network or CXL interfaces can be significant and become harmful to the performance.}
MNs with processing often resemble regular servers ~\cite{disaggmem-nsdi-2017,legoos,disaggmem-eurosys-2018,farmem-nsdi-2014,farmem-eurosys-2020,farmem-osdi-2020,disaggmem-osdi-2020}. Such MNs perform light-weight processing to reduce data movement and network overheads. Given the costs of high-performance CPUs and the light-weight computation required near the data, researchers have proposed MNs that replace the general-purpose processor with light-weight FPGA-based accelerators~\cite{clio,farview}.

In this paper, we perform the first study of disaggregated memory systems for distributed machine learning. We optimize the full system for recommendation, provisioning hardware with CNs and MNs to improve efficiency. We favor MNs with a lightweight CPU/ASIC and show how MN-side computation can support embedding accesses and optimize the full system for recommendation workloads. 

To manage disaggregated CNs and MNs, LegoOS~\cite{legoos} proposed a distributed OS supported by an RDMA-based RPC framework. The RDMA-based network protocol is commonly used by disaggregated memory designs~\cite{farview,clio,disaggmem-asplos-2021,disaggmem-atc-2020,disaggmem-atc-2021} to permit remote memory access without involving host processors. RDMAs can efficiently transfer a large chunk of data across server nodes with speeds approaching that of DRAM memory channels~\cite{nvidia-InfiniBand}

\section{Conclusion}
{\DesName} addresses infrastructure challenges for evolving recommendation systems. It improves system efficiency by independently scaling compute and memory resources to better match changing demands future generations of recommendation models. It also improves system reliability by provisioning backups to handle the different failure rates of CNs and MNs. {\DesName} reduces the total cost of ownership by 49.3\% through disaggregation. 
\revise{
Given the growing trend of increased resource heterogeneity in future datacenters, the flexibility of disaggregation simplifies the deployment of newly added hardware and allows each workload to attain an optimal allocation of hardware resources to maximize the efficiency and the TCO simultaneously. The disaggregated cluster achieves a maximum of 43.6\% cost saving over the monolithic server-based cluster for large-scale, multi-generation recommendation systems.
}

%%%%%%% -- PAPER CONTENT ENDS -- %%%%%%%%

%%%%%%%%% -- BIB STYLE AND FILE -- %%%%%%%%
\bibliographystyle{IEEEtranS}
\bibliography{refs}
%%%%%%%%%%%%%%%%%%%%%%%%%%%%%%%%%%%%

\end{document}